\documentclass[a4paper]{article}

\usepackage{RR}
\usepackage{hyperref}

\usepackage{alltt}
\usepackage{graphicx}
\usepackage{listings}

\RRdate{August 2009}

\RRauthor{
Emmanuel Nataf\thanks{Ma\^itre de conf\'erences - Nancy University}%
  \thanks[sfn]{Madynes INRIA project}%
  \and
Olivier Festor\thanks{Directeur de Recherche INRIA}%
}
\authorhead{E. Nataf, O. Festor}
\RRtitle{jYang : un analyseur yang en java}
\RRetitle{jYang : A \y\ parser in java}
\titlehead{jYang}
\RRresume{Dans  le contexte  du  groupe de  travail administration  et
op\'eration  des r\'eseaux  de l'IETF,  le protocole  de configuration
\nc\   a   \'et\'e   d\'evelopp\'e   pour   manipuler   la   configuration
d'\'equipement   r\'eseau.    \y\  est   le   langage   en  cours   de
standardisation dans  ce groupe. Il permet de  sp\'ecifier des mod\`eles
de donn\'ees  utilisable dans l'approche \nc.  Ce  rapport d\'ecrit la
conception et la r\'ealisation  en java d'un analyseur syntaxique et
s\'emantique de sp\'ecifications \y.}

\RRabstract{The  NETCONF configuration  protocol of  the  IETF Network
Working Group  provides mechanisms to manipulate  the configuration of
network devices.  YANG is  the language currently  under consideration
within the IETF to specify the data  models to be used in \nc\ .  This
report describes the design and  development of a syntax and semantics
parser for \y\ in java.}
\RRmotcle{parser, java, yang, netconf}
\RRkeyword{parser, java, yang, netconf}
 \RRprojet{Madynes}  
 \RRtheme{\THCom} 
 \URLorraine 
 \RCNancy 

\newcommand{\jyang}{{\sl jYang}}
\newcommand{\nc}{NETCONF}
\newcommand{\y}{YANG}
\begin{document}
\makeRT  


\lstset{numbers=left, numberstyle=\tiny, stepnumber=1, numbersep=5pt}
\lstset{escapeinside={(*@}{@*)}}

\section{Introduction}

It is common  in the network management world that  a protocol and a
data model are  separated even if jointly designed,  as it was already
the case in the  SNMP\cite{rfc1157} protocol and its SMI\cite{rfc1155}
data   modeling,CO\-PS\cite{rfc2748}   and  SP\-PI\cite{rfc3159},   or
SMI\-ng\cite{rfc3780} (GDMO and CMIS or  WBEM and CIM outside the IETF
scope).

\nc\ \cite{rfc4741} is the IETF standard that emerged from the netconf
working     group    to     configure     network    devices.      The
netmod\footnote{http://www.ietf.org/html.charters/netmod-charter.html}
working  group defines  \y\ as  a candidate  language to  specify data
models of values  carried by \nc.  This report  describes a YANG parser
called  \jyang\ that provides  a syntaxic  and semantic  validation of
YANG specifications (called modules or sub-modules).

This  report first  provides a  short description  of \nc\  where some
parts  are referenced  by  \y.  Section  \ref{yang}  details the  \y\
language  concepts   and section \ref{api}  details  the   design  and
implementation of the \jyang\ parser.

\section{NETCONF protocol}

\nc\ is a client/server protocol  where the server is a network device
and   the  client   a  management   framework  that   runs  management
applications. Protocol  requests and responses  focus on configuration
manipulation such as getting the current configuration, update, create
or delete all or some part of it. Configurations are represented in an
XML document that contains two sort of data:
\begin{itemize}
\item
configuration  data   that  is  writable   and  that  describes
configuration parameters of the \nc\  agent.
\item
state data that is read-only and that describes operational data such
as counter or statistics.
\end{itemize}

A \nc\ agent can have several configurations each one containing
several configuration data.  There can be only one active
configuration, called the {\tt running} configuration, at the same
time. Other configurations, called {\tt candidate} configurations, can
exist without interfering with the running one.  A special commit
capability (cf section \ref{capabilities}) asks the agent to pass a
candidate configuration as the running one.

\begin{figure}[htbp]
\begin{center}
\includegraphics[scale = .7]{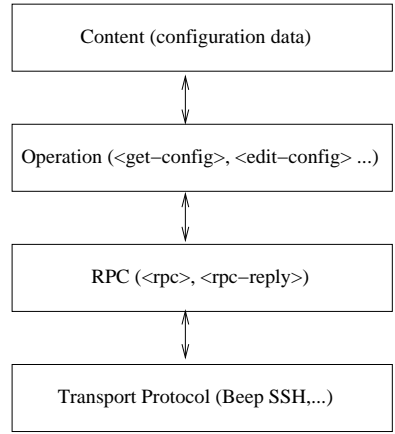}
\end{center}
\caption{NETCONF protocol layers}
\label{netconfarchi}
\end{figure}

Figure  \ref{netconfarchi}  extracted  from \cite{rfc4741}  shows  the
layered protocol architecture of NETCONF.  The protocol mainly defines
operations and how they are carried by rpc mechanisms.

\subsection{Transport protocol}

\nc\  can  use several  connection-oriented  transport protocols.   It
requires  that a  persistent  connection is  maintained between  peers
during a  potentialy long term {\tt  session}.  Ressources reservation
can  be  granted for  the  session  and  any reserved  ressources  are
released at the end of the connection.

Authentication, integrity and confidentiality  must be provided by the
transport  protocol.  A  \nc\  implementation  must  support  the  SSH
transport protocol mapping.

The specification  language described in  this report is not  bound to
the transport protocol used with NETCONF.

\subsection{RPC}

The Remote  Procedure Call  on wich the  \nc\ operations are  built is
described  by two  XML\cite{Bray:00:EML}  elements :  {\tt <rpc>}  for
requests and {\tt <rpc-reply>}  for responses.  The latter can contain
a {\tt <rpc-error>} element when an error occurs during the process of
a request inside the NETCONF agent.

\subsection{Operations}

Basic operations are defined as XML elements :

\begin{itemize}
\item
{\tt <get>} : to retrieve all or part of the running configuration and
state data;
\item
{\tt <get-config>} : to retrieve all or part of a running or candidate
configuration data;
\item
{\tt <edit-config>} : to load all or part of a configuration data to a
specified target running or candidate configuration;
\item
{\tt <copy-config>} : to copy existing configuration data in place of
a specified target running or candidate configuration;
\item
{\tt <delete-config>} : to delete a candidate configuration ;
\item
{\tt <lock>} :  to lock the running configuration  against any edit or
copy  config operations  originated from  another session  or external
access (like SNMP);
\item
{\tt <unlock>} : to unlock a locked configuration;
\item
{\tt  <close-session>}  : to  stop  the  NETCONG  session accepting  any
request but complete operation in progress;
\item
{\tt <kill-session>} : to stop  the NETCONF session without completing any
operation in progress.
\end{itemize}

All operations are  in carried {\tt <rpc>} elements.  A common element
of  get,  edit  and  delete  operations  is  a  filter  element  ({\tt
<filter>})  that   allows  some  filtering   on  data  by   using  the
hierarchical structure of XML documents.

\subsection{Capabilities}
\label{capabilities}

Accepted operations  (basic and news operations) and  data are defined
by capabilities.  A \nc\ agent  can provide more than one capabilities
and  an unique  URI  references each  capabilities.  Capabilities  are
exchanged between entities at session establishment time.

\section{\y}
\label{yang}

The  \y\ Internet-Draft\cite{yang01}  defines \y\  as a  data modeling
language used to describe \nc\  configuration and state data. The \nc\
standard does  not define  such a language  for its content  layer (cf
fig.\ref{netconfarchi}).       The      netmod      working      group
charter\footnote{http://www.ietf.org/html.charters/netmod-charter.html}
explains why  a more hight level  language than XML is  needed (an old
draft           can          be           seen           at          :
http://www.yang-central.org/twiki/\-pub/\-Main/\-Yang\-Do\-cu\-ments/draft-lengyel-why-yang-00.txt).

\subsection{\y\ specifications}

A  \y\ specification contains  formal definitions  of data  types that
will model  real data maintained  by \nc\ agents.   Formal definitions
follow the \y\ syntax. \y\  provides constructs that give semantics to
XML data.  As an XML document  is a collection  of imbricated markups,
\y\   defines  statements   that   can  be   mapped   on  pattern   of
markups.  Moreover  \y\  allows  reusability  of  specifications  with
generic statements or augmentation/extension statements.

\y\ specifications are organized in modules and submodules
that contain data type definitions and operation descriptions. 

\subsection{\y\ module and submodule headers}

\y\  modules  and submodules  have  some  headers  that are  informations
related to the module or submodule itself.

\subsubsection{Module header}
\label{spec:module}

A module has mandatory headers and one optional header. The mandatory
ones are the {\tt name space} and {\tt prefix}.  For example :

\begin{lstlisting}
module router {
  namespace ``urn:madynes:xml:ns:yang:router'';(*@\label{namespace}@*)
  prefix router;(*@\label{prefix}@*)
...
\end{lstlisting}

The name space at line \ref{namespace}  is for all data defined in the
module  and the prefix  at line  \ref{prefix} can  be used  inside the
module (when  confusion is  possible) to refer  some data. A  \y\ {\tt
version} header is optional.

\subsubsection{Submodule header}
\label{spec:submodule}

A submodule has one or two headers. It must have a {\tt belongs
to} statement and may have the \y\ {\tt version} statement. A
submodule belongs to one and only one module. For example :

\begin{lstlisting}
submodule routing-policies {
  belongs-to router ;(*@\label{belongsto}@*)

...
\end{lstlisting}

The  submodule  {\tt routing-policies}  belongs  to  the {\tt  router}
module at line \ref{belongsto}.

\subsubsection{Yang specification meta statements}

Meta  statements  give some  general  information  on  the module  or
submodule.  These  informations  concern  the  organization  that
defines the module, the contact,  the description and the reference of
the  \y\ specification.   At most four meta statements can be made.
A meta statement of a  specification must not be duplicated (e.g. two
contact meta statement in a module).

\subsubsection{Yang linkage statements}

A  yang  specification  can   have  {\tt  import}  and  {\tt  include}
statements.

\paragraph{Import statement}

The syntax  allows to identify another module and associate  it to a
prefix. For example :

\begin{lstlisting}
module router {
...
  import yang-types {  (*@\label{import}@*)
    prefix yang;       (*@\label{prefiximp}@*)
  }
...
\end{lstlisting}

The module {\tt  yang-types} is imported at line  \ref{import} so that
any  type or  data defined  in this  module can  be used  in  the {\tt
router} module. In order to use them without conflict, the prefix {\tt
yang} defined at line \ref{prefiximp} must be used. For example (again
in the {\tt router} module):

\begin{lstlisting}
...
leaf network {
  type yang:counter32;
}
...
\end{lstlisting}

where {\tt counter32}  is defined in the {\tt  yang-types} module. The
prefix  used must  be the  same than  the one  defined in  the prefix
statement of the imported  module (see section \ref{spec:module}). 

 There can be several import statements but each prefix must be unique
in the  module.  The prefix  defined in a  module can be used  in this
module. A submodule can import modules but no submodules.

\paragraph{Include statement}
\label{spec:include}

The syntax allows to refer to a submodule. For example:

\begin{lstlisting}
module router {
...
include routing-policies; (*@\label{include}@*)
...
\end{lstlisting}

The {\tt router} module  includes the {\tt routing-policies} submodule
at line \ref{include} so any type or data defined in the submodule can
be used in that {\tt router} module.

An included submodule must have  a {\tt belongs-to} statement with the
reference of the  including module (see section \ref{spec:submodule}).
A submodule can  include other submodules but they  must all belong to
the same module.

\subsubsection{Yang revision statement}

Any yang specification should  contain revision statements.  There is
one YANG\_Re\-vi\-si\-on  instance for each yang  revision statement and
each one can contain none or one description statement.

\y\ specifications  describe1 data as a  tree of nodes.  There are two
main node  types; {\bf leaf} nodes  that contain data  values and {\bf
construct}  nodes that  contain  (in the  hierarchical meaning)  other
nodes.

\subsection{Leaf nodes}

There are two classes of leaf nodes~:
\begin{itemize}
\item
({\tt leaf}) that contains one value;
\item
({\tt leaf-list}) that contains a list of values of the same
type.
\end{itemize}

\subsection{Construct nodes}

A construct node definition  contains other node definitions. Value of
such a node depends on the type of the construct node:
\begin{itemize}
\item
{\tt container} that contains other nodes and its value is composed of
values of all contained nodes;
\item
{\tt list} that contains other nodes and its value is composed of
several values of all contained nodes.  A list value can be seen  as a two
dimensional array  and a {\tt key}  parameter of the  {\tt list} allows
the reference of one instance of the list of node (an entry);
\item
{\tt choice} that defines {\tt case} constructs containing other nodes
and its value is the value of contained nodes of one of the defined
cases;
\item
{\tt rpc} that  contains other nodes and is used  in the rpc mechanism
of \nc and its value is the value of contained nodes.
\item
{\tt  notification} that  contains other  nodes  and is  used by  \nc\
notifications and its value is the value of contained nodes.
\end{itemize}

Following is an example of  a part of a \y\ specification\footnote{All
example in  this report are  inspired from the  draft\cite{yang01}} that
describes  a table  of network  interfaces, a  conceptual view  of two
entries and the XML document of this configuration:

\noindent
\begin{tabular}{lcl}
\begin{minipage}{.25\textwidth}
\begin{verbatim}
list interfaces {
   key index;
   leaf index {
     type int8;
   }
   leaf name {
     type string;
   }
   leaf type {
     type string;
   }
   leaf speed {
     type int64;
   }
}
\end{verbatim}
\end{minipage}
&
\begin{minipage}{.55\textwidth}
\includegraphics[scale=.6]{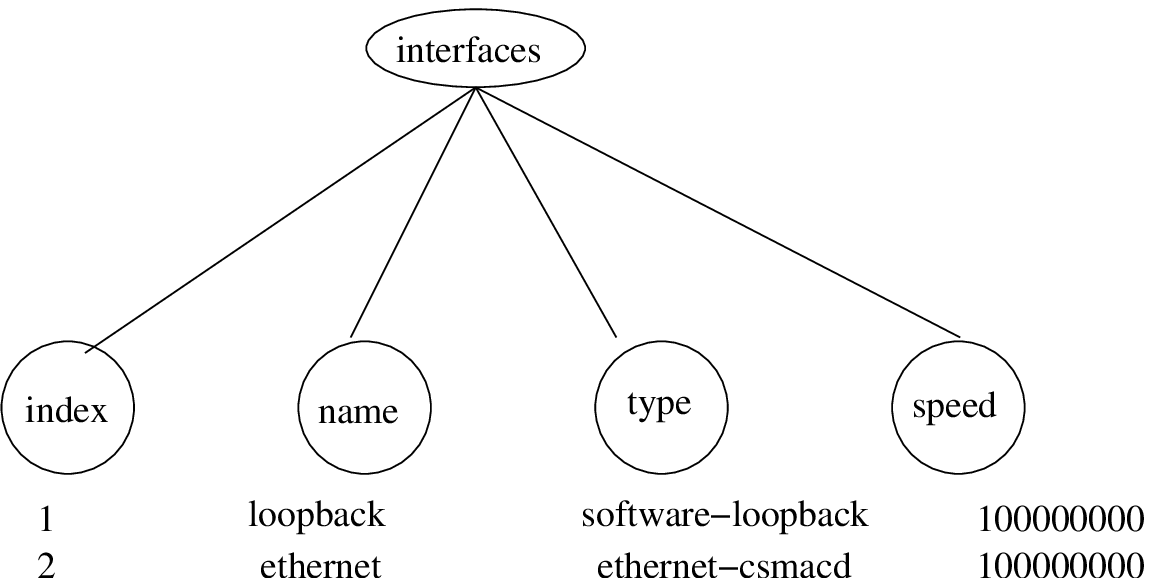}
\end{minipage}
&
\begin{minipage}{.2\textwidth}
\begin{small}
\begin{verbatim}
<list>
  <index>
     1
  </index>
  <name>
     loopback
  </name>
  <type>
     software-loopback
  </type>
  <speed>
     100000000
  </speed>
  <index>
     2
  </index>
  <name>
     ethernet
  </name>
  <type>
     ethernet-csmacd
  </type>
  <speed>
     100000000
  </speed>
</list>
\end{verbatim}
\end{small}
\end{minipage}
\end{tabular}

\subsection{Typedef}

\y\ defines  a set  of base types  (integer, float,  string\ldots) and
allows  the definition  of  new types  from  existing ones  by a  {\tt
typedef} construct.  For example below  is the definition of a 32 bits
counter from the basic unsigned integer {\tt uint32}.

\begin{lstlisting}
 typedef counter32 {
        type uint32;
        description
           "The counter32 type represents...
        reference
           "RFC 2578 (STD 58)";
    }
\end{lstlisting}

New  types can  be used  in  data nodes  and in  other {\tt  typedef}.
Depending on the base type  used in a {\tt typedef}, some restrictions
can be  added like  a range  restriction on numerical  values or  as a
string  pattern on  string derived  types. When  defining a  new type,
restrictions must  only restrict the value  set of the  base type. The
new type is a sub-type of the base type.

\subsection{Grouping and Uses}

\y\ provides a reusability concept  with {\tt grouping} and {\tt uses}
statements. A  grouping is a  set of definitions (leafs  and construct
nodes, typedef, grouping\ldots) that  can be used in other definitions
with the {\tt uses} statement.  For example below is the definition of
the grouping {\tt  address} with two leaf nodes  at lines \ref{leafip}
and  \ref{leafport}   and  its  usage  in   the  {\tt  http-container}
container (line \ref{leafhttpserv}).

\noindent
\begin{tabular}{lr}
\begin{minipage}{.5\textwidth}
\begin{lstlisting}[name=grouping]
grouping address {
    leaf ip { (*@\label{leafip}@*)
      type bits (32);
    }
    leaf port {  (*@\label{leafport}@*)
      type uint32;
    }
}
\end{lstlisting}
\end{minipage}
&
\begin{minipage}{.5\textwidth}
\begin{lstlisting}[name=grouping]

container http-server { (*@\label{leafhttpserv}@*)
  leaf name { 
    type string; 
  }
  uses address;
}
\end{lstlisting}
\end{minipage}
\end{tabular}

This construct is equivalent to :

\noindent
\begin{lstlisting}

container http-server {
  leaf name { 
    type string; 
  }
  leaf ip {
    type bits (32);
  }
  leaf port {
    type uint32;
  }
}
\end{lstlisting}

\subsection{Augmenting}

The {\tt augment}  statement contains nodes and is  used to add theses
nodes to  an existing  construct node. In  the specification  below, a
container named  {\tt login} at  line \ref{logincont} contains  a leaf
named {\tt  message} line  \ref{leafmess} and a  list {\tt  user} line
\ref{listuser} having  several leaf nodes (just {\tt  name} is shown).
The  {\tt augment} statement  at line  \ref{augmentuse} refers  to the
list {\tt  user} under the  container {\tt login}  and adds to  it the
leaf {\tt uid} at line \ref{uid}.

\noindent
\begin{tabular}{lr}
\begin{minipage}{.5\textwidth}
\begin{lstlisting}[name=augment]

  container login { (*@\label{logincont}@*)
    leaf message {  (*@\label{leafmess}@*)
      type string;
      }
    list user {      (*@\label{listuser}@*)
      key ``name'';
      leaf name { 
         type string;
      }
      ...
    }
\end{lstlisting}
\end{minipage}
&
\begin{minipage}{.5\textwidth}
\begin{lstlisting}[name=augment]
augment login/user {  (*@\label{augmentuse}@*)
  leaf uid {           (*@\label{uid}@*) 
    type uint16;
  }
}
\end{lstlisting}
\end{minipage}
\end{tabular}

Note that augmenting is not the  same as grouping. Grouping is used to
reduce the  size of  a specification by  using several times  the same
construct  while  augmenting  allows  to  add  nodes  to  an  existing
one.  Augmenting  is useful  when  an  equimement has  vendor-specific
parameters added to standard ones.

\subsection{Rpc}

As  a  \nc\  agent  can  provide capabilities  with  new  rpc  embeded
operations,  \y\ allows the  specification of  such an  operation. For
example  the  {\tt  activate-software}  operation below  defines  data
sended  in a  {\tt <rpc>}  message  with {\tt  input} statement  (line
\ref{input}) and  data returned in  a {\tt <rpc-reply>} with  the {\tt
ouput} statement (line \ref{output}).

\begin{lstlisting}
rpc activate-software-image {
    input {(*@\label{input}@*)
       leaf image-name {
         type string;
       }
    }
    output {(*@\label{output}@*)
       leaf status {
         type string;
       }
    }
}
\end{lstlisting}

\subsection{Notification}

A \nc\ agent can send notifications  that can be specified with \y\ by
the  {\tt   notification}  statement.   Nodes  contained   in  a  {\tt
specification} statement  model data sent  by the agent.  Below  is an
example where the index  of a failed interface (line \ref{notifparam})
will be sent.

\begin{lstlisting} 
notification link-failure {
  description "A link failure has been detected";
  leaf if-index {(*@\label{notifparam}@*)
      type int32 { range "1 .. max"; }
  }
}
\end{lstlisting}

\subsection{Extensions}

\y\ allows  the definition of  new statements when  specific processes
requires  it. The  content of  an extention  is to  be  interpreted by
specific  implementation.   Extensions can  be  used  anywhere in  \y\
specifications. In the example  below, the extension {\tt c-define} is
specified and  used with one name  argument (line \ref{extensionarg}).
Each use of  an extension must be prefixed by  the module prefix where
the extension is defined.

\begin{lstlisting}
extension c-define {
         description
           "Takes as argument a name string.
           Makes the code generator use the given name in the
           #define.";
         argument "name";(*@\label{extensionarg}@*)
       }
\end{lstlisting}

\begin{lstlisting}
 myext:c-define "MY_INTERFACES";
\end{lstlisting}

\subsection{YIN}

YIN  is an alternative  XML-based syntax  for \y\  specifications. YIN
specifications can be generated from  \y\ ones and are equivalent. The
goal of YIN specifications is to enable seemless interactions with XML
based tools  (as XSLT).  \jyang\  parser allows the generation  of YIN
specifications from \y.

\section{\jyang}

\jyang\ is  a java  parser for \y\  specifications and  an application
programming interface  offering a programmatic  access in java  to \y\
specifications.

\subsection{\y\ Parser}

The    java    parser    is    built   with    JJTree    and    JavaCC
\footnote{https://javacc.dev.java.net}  but  no  external  library  is
needed to use it.

\begin{itemize}
\item
lexical and syntax  checks are conformant to the  ABNF grammar given in
\cite{yang01}
\item
semantical check covers following features :
\begin{itemize}
\item
name scoping and accessibility for typedef, grouping, extension, uses,
leaf and  leaflist, inside  a module  or submodule  and with  imported and
included specifications.
\item
type restriction  for any type (integer,  boolean, bits, float,\ldots)
and typedef
\item
default value and restriction
\item
augment existing node
\item
Xpath for schema node in augment, leaf (of key ref type) and list (for
unique statement)
\end{itemize}
\end{itemize}

\subsection{Repository}

\jyang\ is  an open source distribution  of our toolkit  under the GPL
licence. The official repository is at the INRIA Gforge web site :\\
\begin{verbatim}
http://jyang.gforge.inria.fr
\end{verbatim}

\subsection{\jyang\ tools}

\subsubsection{\jyang\ parser use}

\jyang\ is distributed  as a java jar file  called {\tt jyang.jar} and
configured to be executable. The synoptic is :

\begin{verbatim}
java -jar jyang.jar [-h] [-f format] [-o outputfile] [-p paths] file [file]*
\end{verbatim}

\begin{itemize}
\item
{\tt -h} print the synoptic
\item
{\tt  -f format}  specifies the  format for  a translated  output (yin
format for example)
\item
{\tt -o outputfile} the name of the translated output (standard output
if not given) ignored if no format are given
\item
{\tt -p  paths} a path where  to find other \y\  specifications. It is
needed   if  import  or   include  statements   are  in   the  checked
specification or  if the environement variable {\tt  YANG\_PATH} is not
set.
\item
{\tt file  [file]*} specifies  files containing \y\  specification. It
must be  one specification ({\tt  module} or {\tt submodule}  for each
file).
\end{itemize}

\paragraph{Errors}

Errors in \y\ specifications are printed on the standard error output.
\jyang\ stops  checking at the  first lexical or syntaxical  error but
tring to  check after  a first semantical  error is  encountered. When
such an error  is detected, the current bloc  statement is escaped and
\jyang\ passes to the next statement.

\subsubsection{Programmatic access}

\jyang\   provides  java   classes   and  interfaces   to  parse   \y\
specification inside a java  program. Internal representation of those
specifications can be accessed throught the API defined in the section
\ref{api}. Below is an example of how to parse a \y\ specification.

\begin{lstlisting}
import java.io.*;
import jyang.*;

public class JyangTest {

   /**
   * Simple jyang test, parses and checks one YANG specification.
   * Imported or included modules or submodules are looked in the 
   * current directory.
   * Error messages are on the standard output
   * 
   * @param args YANG file name
   */
   public static void main(String[] args) throws Exception {
        FileInputStream yangfile = new FileInputStream(args[0]);(*@\label{getyangfile}@*)
        new yang(yangfile);(*@\label{jyangparser}@*)
        YANG_Specification spec = yang.Start();(*@\label{startjyang}@*)
        spec.check();(*@\label{jyangcheck}@*)
   }
}
\end{lstlisting}

The   program  first  gets   the  \y\   specification  file   at  line
\ref{getyangfile}.    A    new   jyang   parser    is   created   line
\ref{jyangparser} with this file.  The lexical and syntactic check are
processed  at line  \ref{startjyang} and  return  a YANG\_specification
object  instance  that  can   be  semantically  checked,  as  at  line
\ref{jyangcheck}.

\section{\jyang\ API}
\label{api}

\subsection{UML class diagram}

Following  sections contain  the  UML class  diagrams  of the  \jyang\
API. UML classes (abstacts or not) are java classes and UML interfaces
are java interfaces.  Inheritance relations are directly mapped to the
java  inheritance mechanism (we  have limited  in the  design multiple
inheritance to interfaces only).

For relationships other than inheritance the API follows theses rules :
\begin{itemize}
\item
when  the cardinality  is {\tt  0-1} there  is a  getter and  a setter
method  with  the name  of  the related  class  in  the other  related
class. For example in figure  \ref{extension} there is a method called
{\tt  getArgument} in the  {\tt YANG\_Extension}  java class  and this
method  returns   an  instance   of  the  {\tt   YANG\_Argument}  java
class. Such method returns {\tt  null} if there is no related instance
(but some relations have no {\tt 0} lower bound and so must not return
null).      There     is     also     a     method     called     {\tt
setArgument(YANG\_Argument)}.
\item
when  the cardinality  is {\tt  0-n} the  getter returns  a  java {\tt
Vector} instance containing related instances. The getter has an extra
's', for  example in  the figure \ref{spec}  there is a  method called
{\tt getLinkages()}  in the  {\tt YANG\_Specification} java  class. If
there is  no related instance, the  method returns an  empty java {\tt
Vector}. For the setter, as it  is often used during parsing, there is
a  method   called  {\tt   add}{\sl  Class-Name}  (for   example  {\tt
addLinkage(YANG\_Linkage}).
\end{itemize}
 
\subsection{\y\ specifications}

\begin{figure}[htbp]
\begin{center}
\includegraphics[scale = .3]{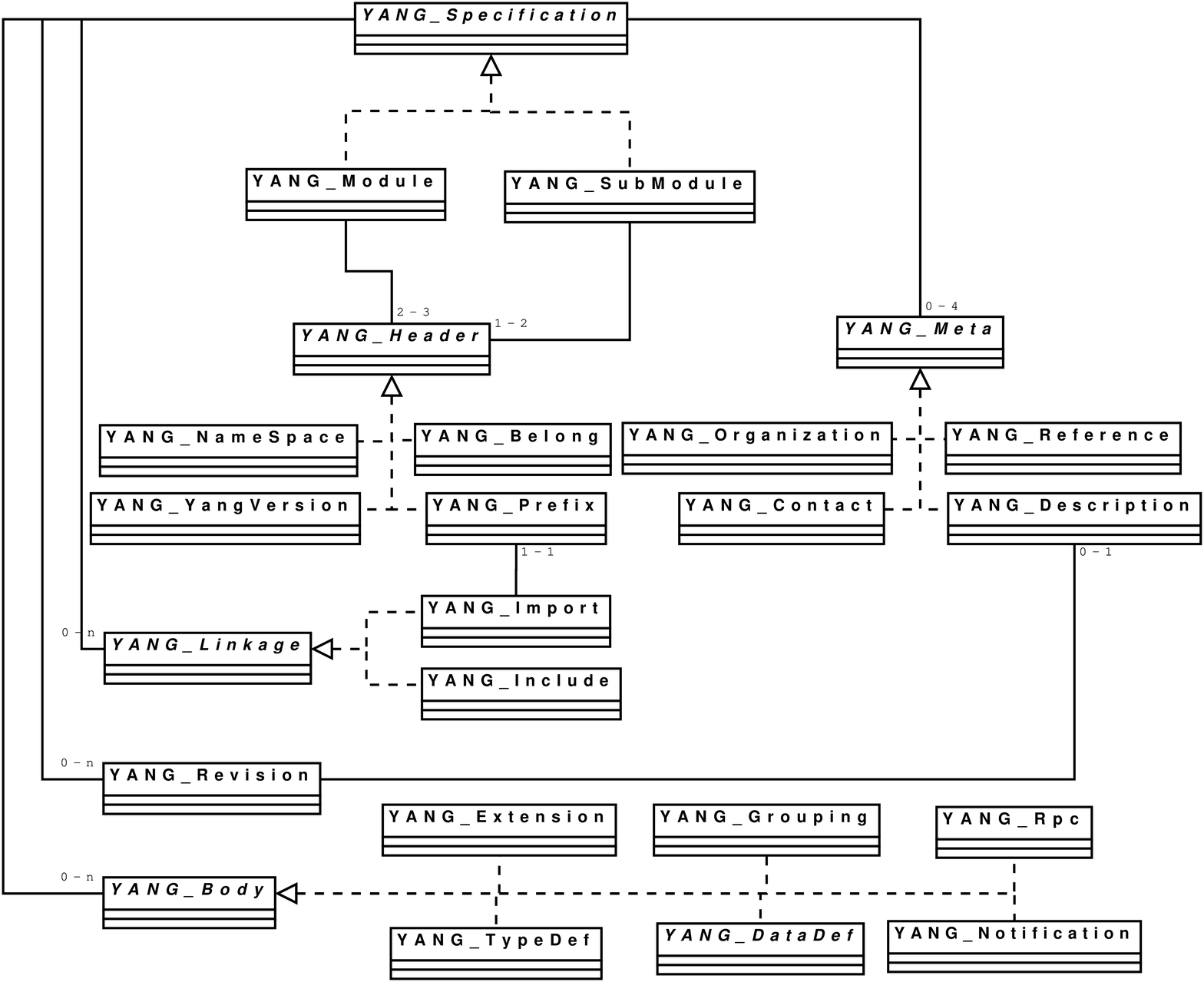}
\end{center}
\caption{Module and SubModule}
\label{spec}
\end{figure}

Figure  \ref{spec}   shows  the  top  level   classes  and  interfaces
hierarchy.  On top is the  YANG\_Specification interface that can be a
YANG\_Module  for  a  yang  module  or a  YANG\_SubModule  for  a  \y\
submodule.

\subsection{Yang body statements}

Data definitions are  in body statements that can  be: extension, type
definition,  grouping,  data  definition,  rpc or  notification.   The
YANG\_Body interface is the common  interface for all bodies in a yang
specification.

\subsection{Bodies}

\subsubsection{Extension statement}
\label{extension:section:global}

\begin{figure}[htbp]
\begin{center}
\includegraphics[scale = .3]{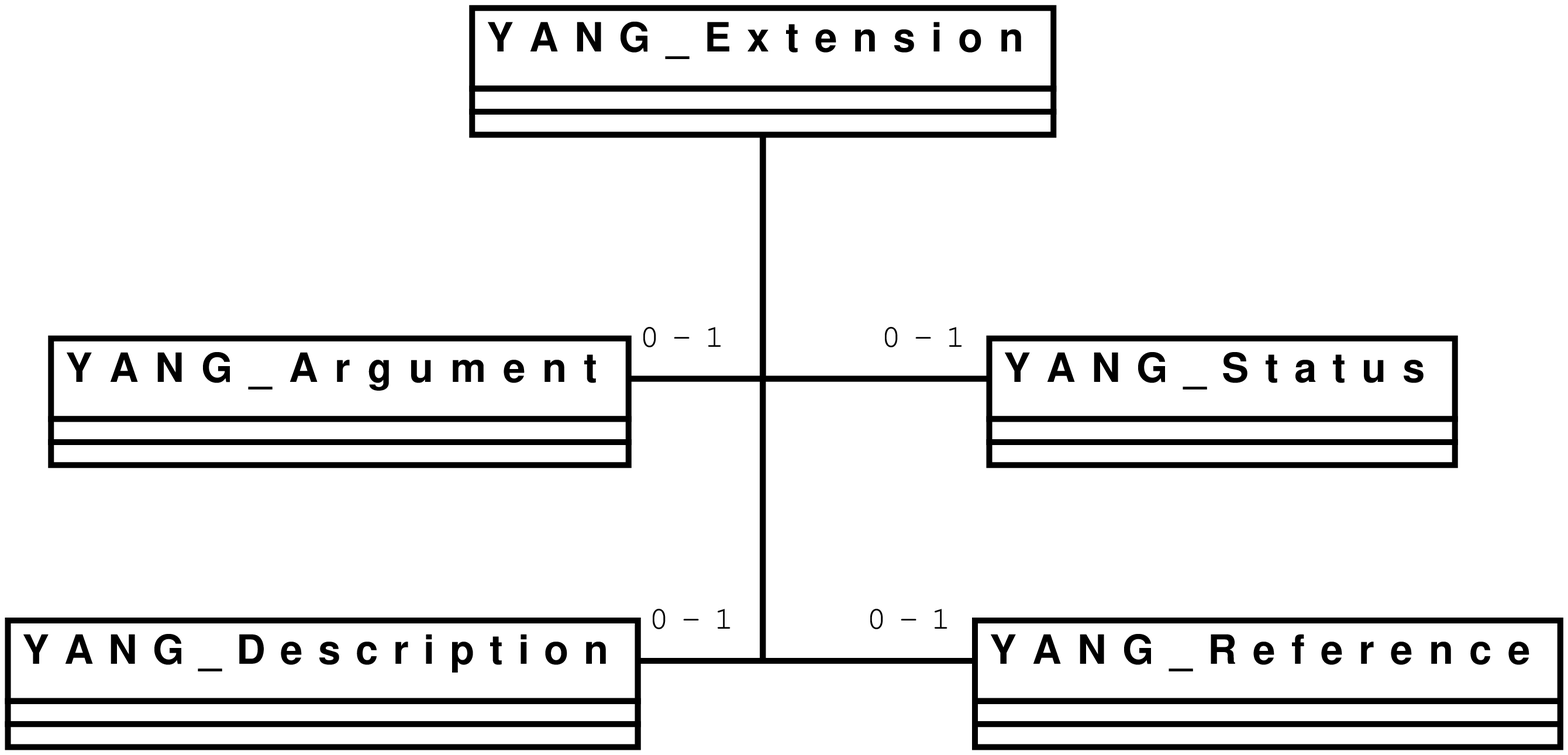}
\end{center}
\caption{Extension statement classes}
\label{extension}
\end{figure}

An extension  statement (fig.  \ref{extension}) can be  stand alone or
can contain  other statements either as  argument, status, description
and  reference.  Each  of these  statements  can occur  at most  once.
Their       description       is       detailed       in       section
\ref{extension:section:detail}.

\subsubsection{TypeDef statement}
\label{typedef:section:global}

\begin{figure}[htbp]
\begin{center}
\includegraphics[scale = .3]{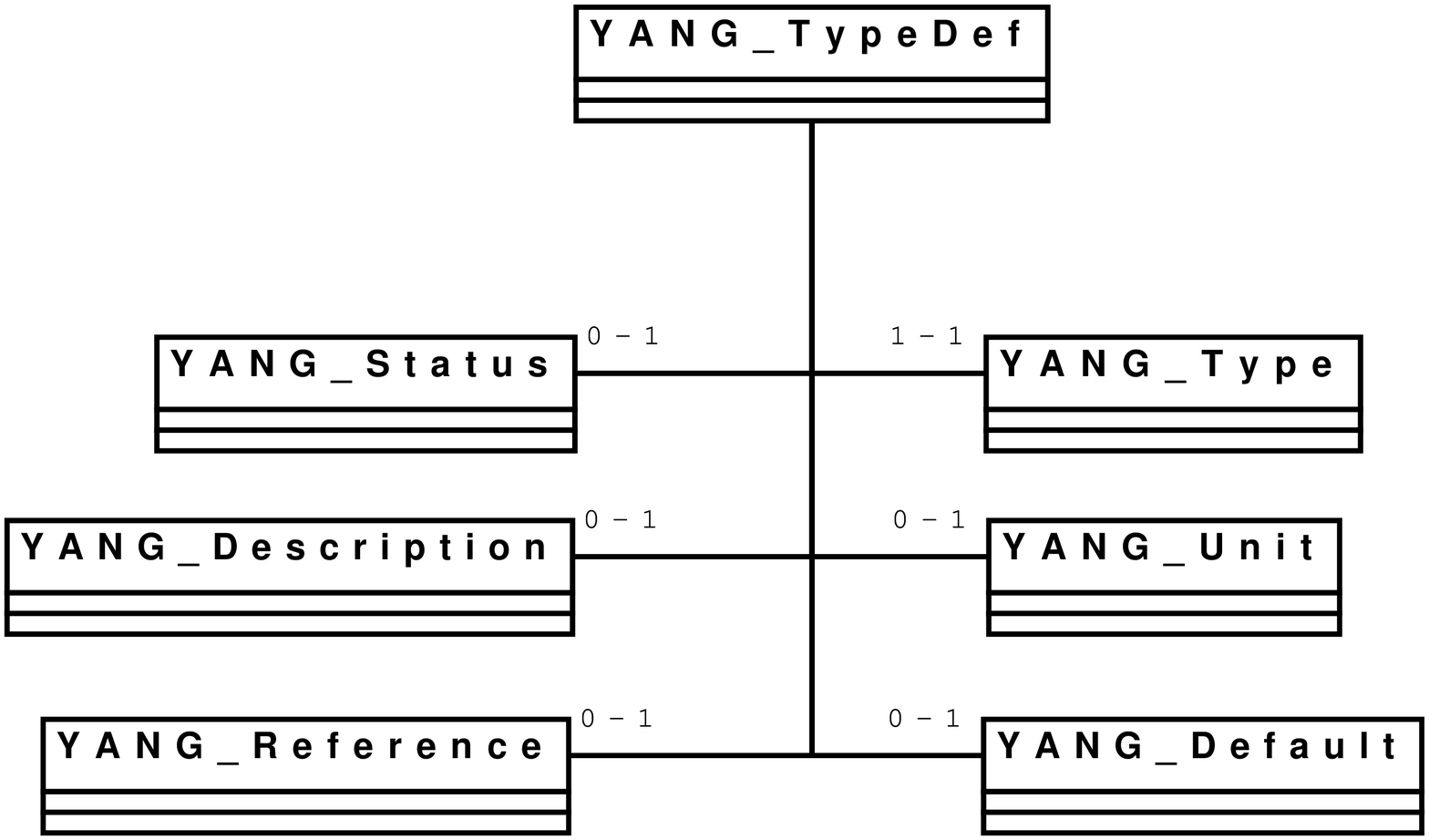}
\end{center}
\caption{TypeDef statement classes}
\label{typedef}
\end{figure}

A  typedef  statement  (fig.    \ref{typedef})  must  contain  a  type
statement  and can  contain  units, default,  status, description  and
reference  statements.  Each  of these  statements can  occur  at most
once.      Their     description     is    detailed     in     section
\ref{typedef:section:detail}.

\subsubsection{Grouping statement}
\label{grouping:section:global}

\begin{figure}[htbp]
\begin{center}
\includegraphics[scale = .3]{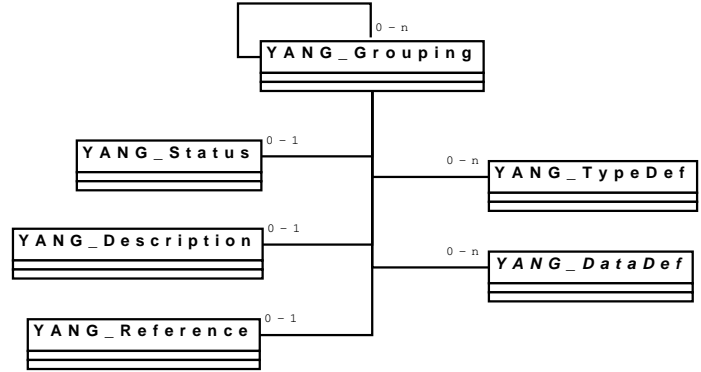}
\end{center}
\caption{Grouping statement classes}
\label{grouping}
\end{figure}

A grouping statement (fig.  \ref{grouping})  can be single or can
contain status,  description and reference statements.   Each of these
statements can occur at most one time. A grouping statement can also
contain  several  other  grouping,  typedef and  datadef  statements.
Their       description       is       detailed      in       section
\ref{grouping:section:details}.

\subsubsection{DataDef statement}
\label{datadef:section:global}

\begin{figure}[htbp]
\begin{center}
\includegraphics[scale = .3]{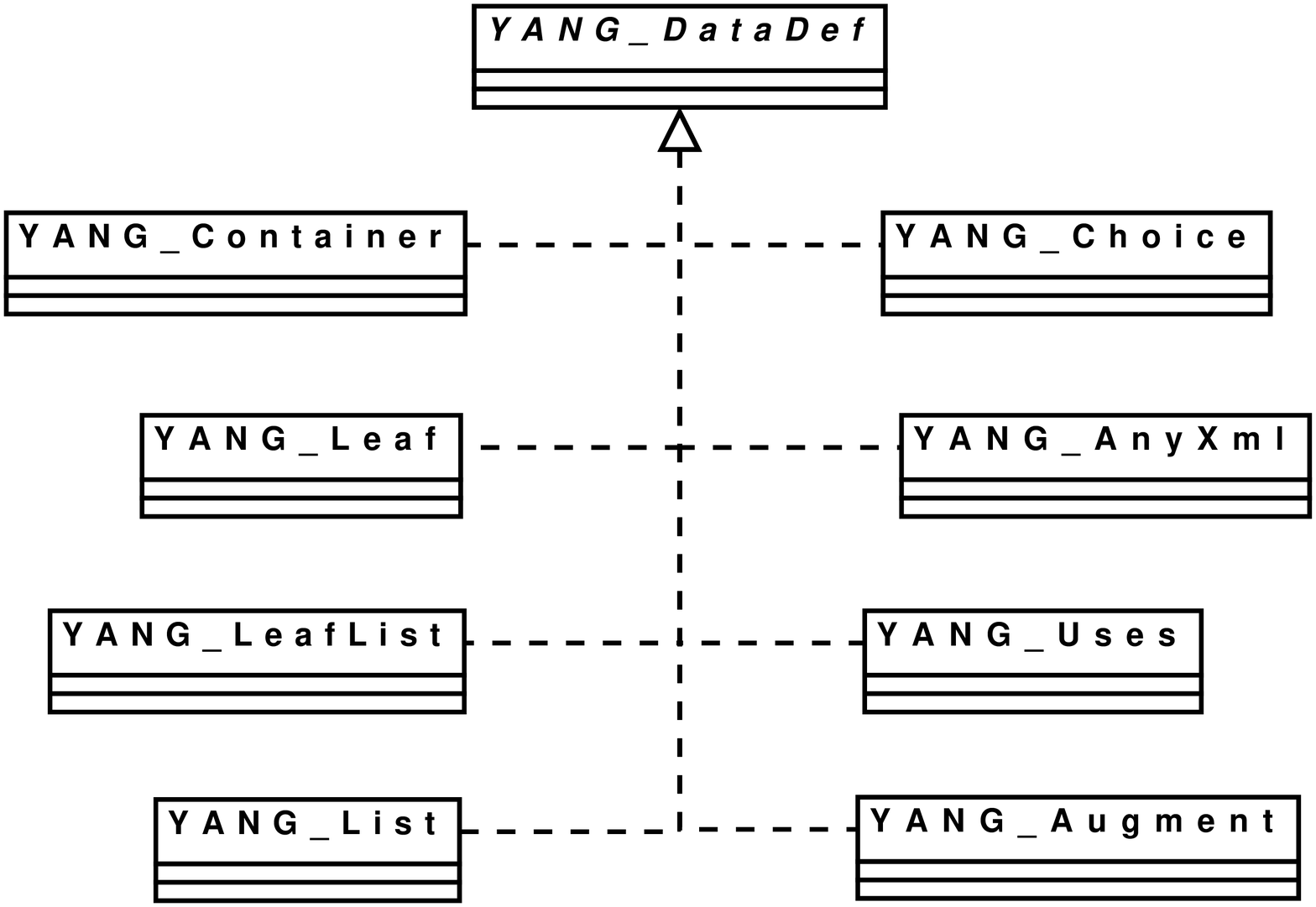}
\end{center}
\caption{DataDef statement classes}
\label{datadef}
\end{figure}

A datadef  statement (fig. \ref{datadef}) is either  a leaf, leaflist,
list, choice,  anyxml, uses or augment  statement. Their description is
detailed in section \ref{datadef:section:detail}.

\subsubsection{Rpc statement}

\begin{figure}[htbp]
\begin{center}
\includegraphics[scale = .3]{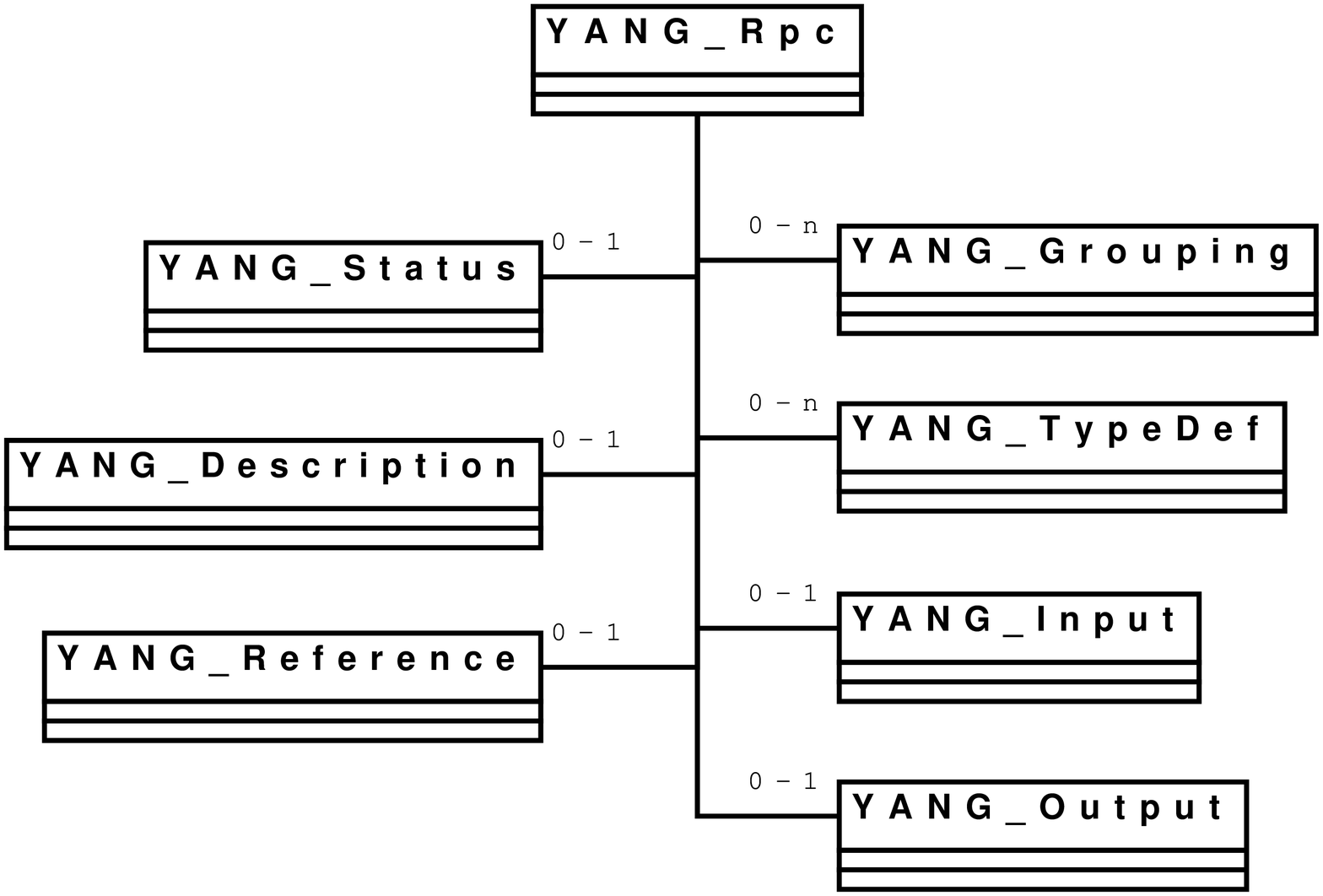}
\end{center}
\caption{Rpc statement classes}
\label{rpc}
\end{figure}

A rpc statement (fig.  \ref{rpc})  can be alone or can contain status,
description, reference,  input and  output statements.  Each  of these
statements can occur  at most once.  A rpc  statement can also contain
several other grouping, typedef and datadef statements.

\subsubsection{Notification statement}

\begin{figure}[htbp]
\begin{center}
\includegraphics[scale = .3]{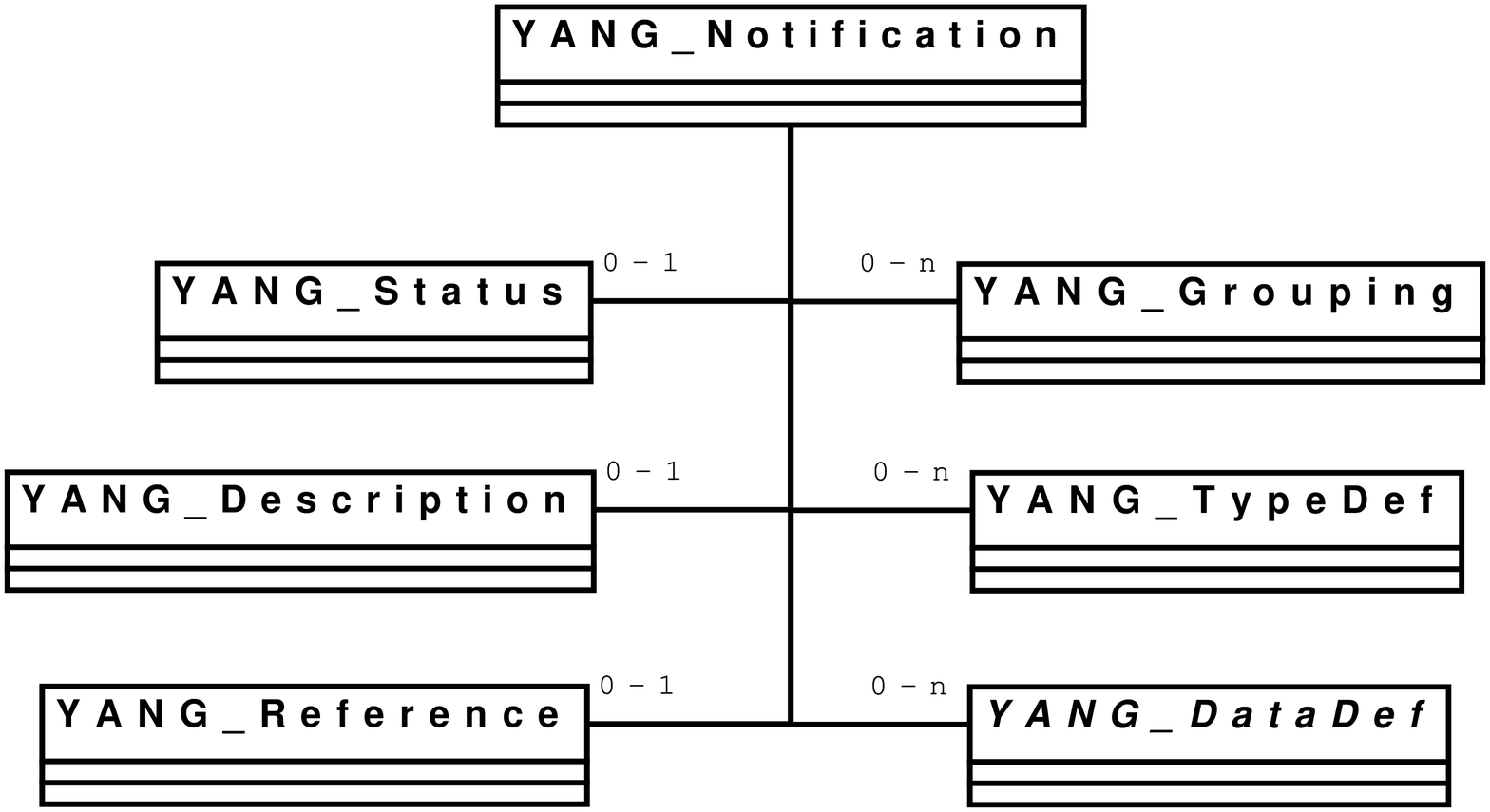}
\end{center}
\caption{Notification statement classes}
\label{notification}
\end{figure}

A notification  statement (fig.   \ref{notification}) can be  alone or
can  contain status,  description and  reference statements.   Each of
these statements can occur at  most once. A notification statement can
also contain several other grouping, typedef and datadef statements.

\subsection{Extension details}
\label{extension:section:detail}

This section refers  to the section \ref{extension:section:global}. It
details all statements that can occur in an extension statement.

\subsubsection{Argument statement}

\begin{figure}[htbp]
\begin{center}
\includegraphics[scale = .3]{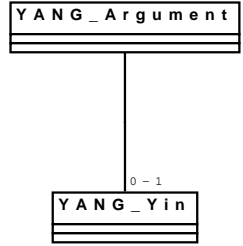}
\end{center}
\caption{Argument statement classes}
\label{argument}
\end{figure}

An  argument (fig.   \ref{argument}) is  composed of  at most  one yin
statement.  A  yin  statement  contains  either the  ``true''  or  the
``false'' string.

There is no more syntax checking needed by other extension
substatements (description, status and reference).

\subsection{Typedef detail}
\label{typedef:section:detail}

This  section refers to  the section  \ref{typedef:section:global}. It
details all statements that can occur in a typedef statement.

\subsubsection{Type statement}
\label{type:section:global}

\begin{figure}[htbp]
\begin{center}
\includegraphics[scale = .3]{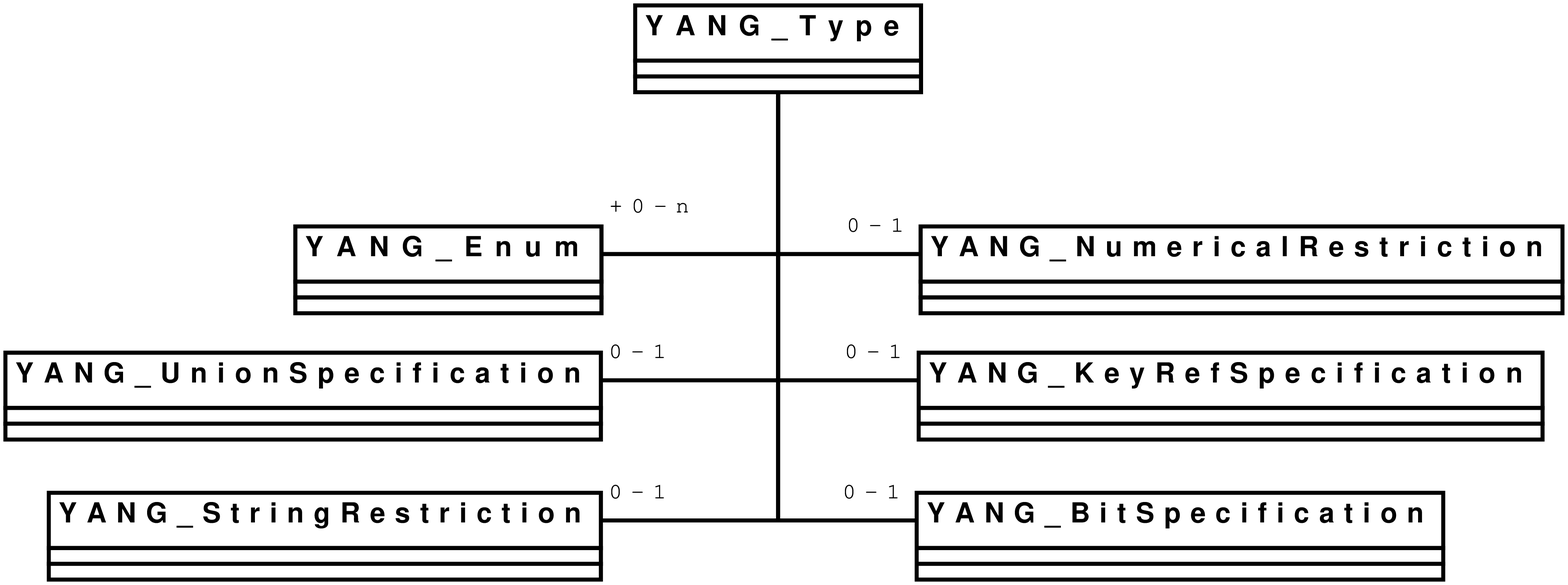}
\end{center}
\caption{Type statement classes}
\label{type}
\end{figure}

 A  type (fig.  \ref{type}) is  composed of  either one  or  more enum
 statement or only one of the specification or restriction statement.

There  is   no  more  syntax   checking  needed  by   other  typedef
substatements  (description, status, default  and units).  Default and
units statements are subject to semantical checking.

\subsection{Grouping detail}
\label{grouping:section:details}

This section  refers to the  section \ref{grouping:section:global}. It
does not  detail any statement  like status, description  and reference.
Typedef is detailed  in the section \ref{typedef:section:detail}.  The
{\tt    data-def}   statements   are    detailed   in    the   section
\ref{datadef:section:detail}.

\subsection{Data def details}
\label{datadef:section:detail}

This  section refers to  the section  \ref{datadef:section:global}. It
details those statements that can be a {\tt data-def} statement.

\subsubsection{Container statement}

\begin{figure}[htbp]
\begin{center}
\includegraphics[scale = .3]{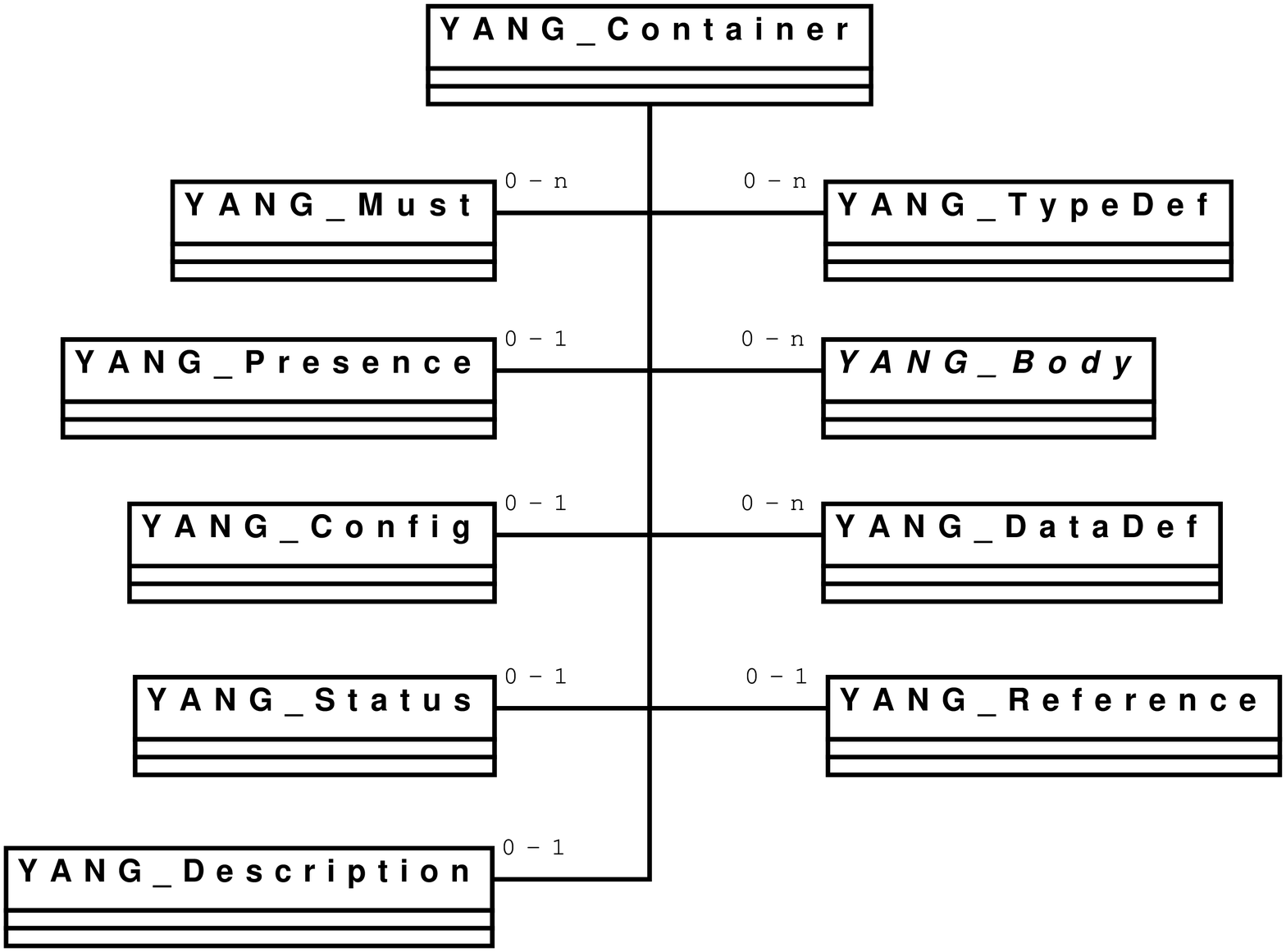}
\end{center}
\caption{Container statement classes}
\label{container}
\end{figure}

A  {\tt  container}  statement  (fig.   \ref{container})  can  contain
several  must, typedef,  grouping and  data-def  statements. Presence,
config, status, description and reference statements are optional.

\subsubsection{Leaf statement}

\begin{figure}[htbp]
\begin{center}
\includegraphics[scale = .3]{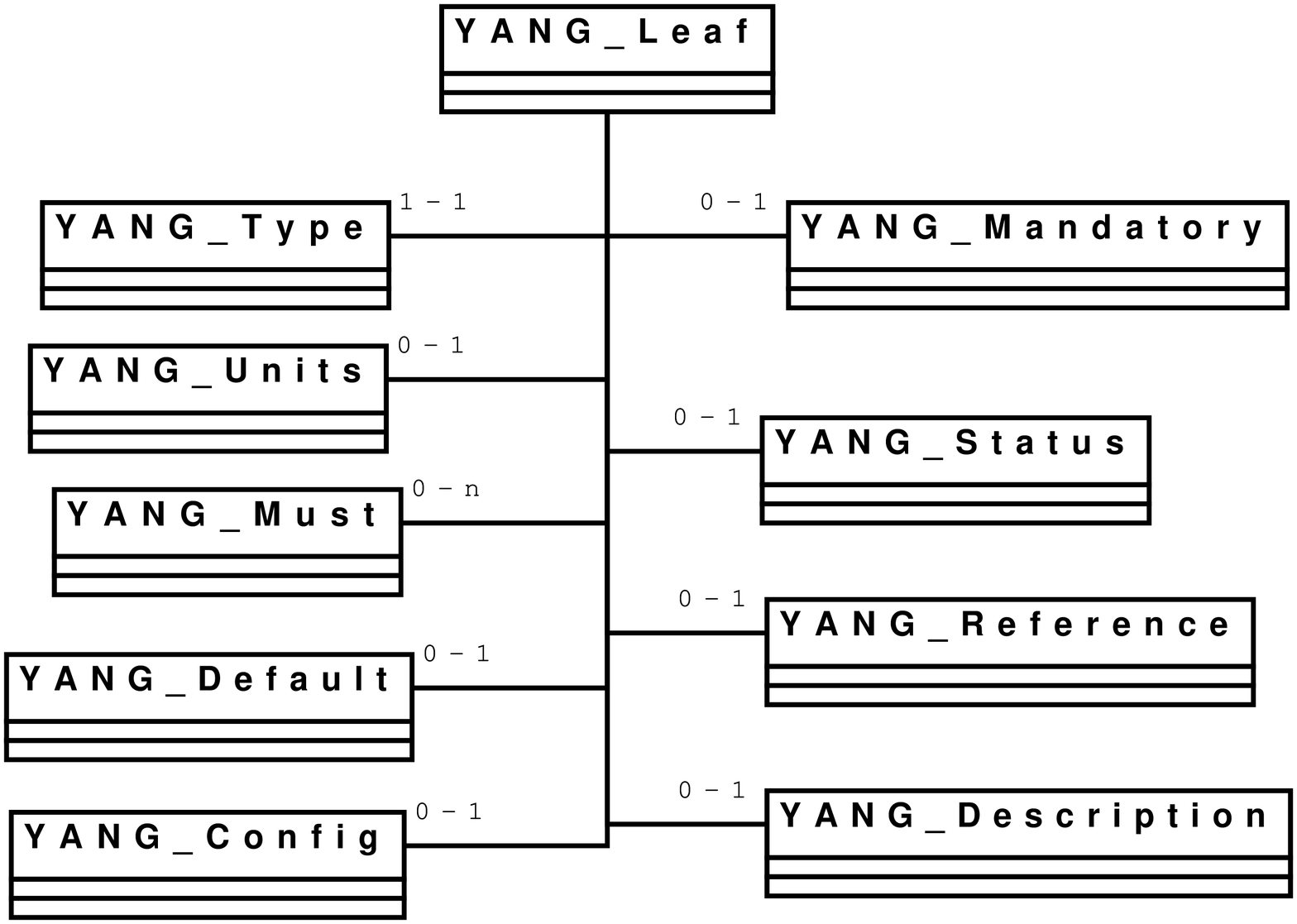}
\end{center}
\caption{Leaf statement classes}
\label{leaf}
\end{figure}

A  {\tt  leaf}  statement  (fig.  \ref{leaf}) must  contain  one  type
statement  (see section  \ref{type:section:global})  and several  must
statements.  Units, default,  config, mandatory, status, reference and
description are optional.

\subsubsection{Leaf List statement}

\begin{figure}[htbp]
\begin{center}
\includegraphics[scale = .3]{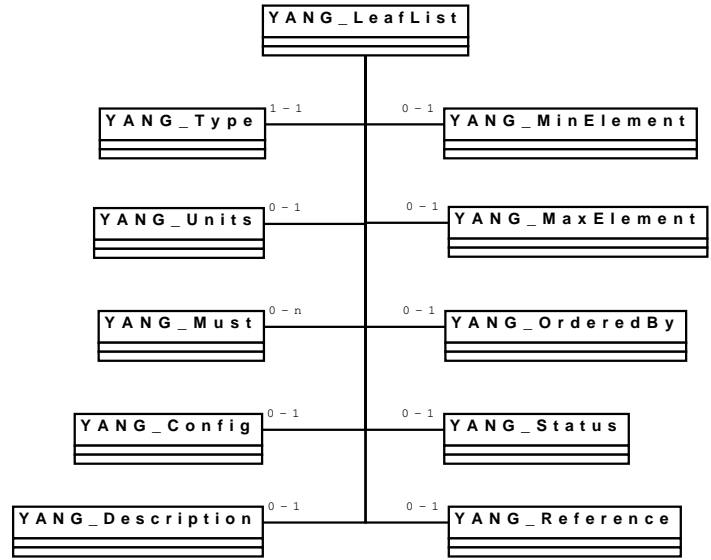}
\end{center}
\caption{Leaf list statement classes}
\label{leaflist}
\end{figure}

A {\tt  leaf-list} statement  (fig.  \ref{leaflist}) must  contain one
type statement  (see section \ref{type:section:global}),  several must
statements.   Units,  default,   config,  min  element,  max  element,
mandatory, status, reference and description are optional.

\subsubsection{List statement}

\begin{figure}[htbp]
\begin{center}
\includegraphics[scale = .3]{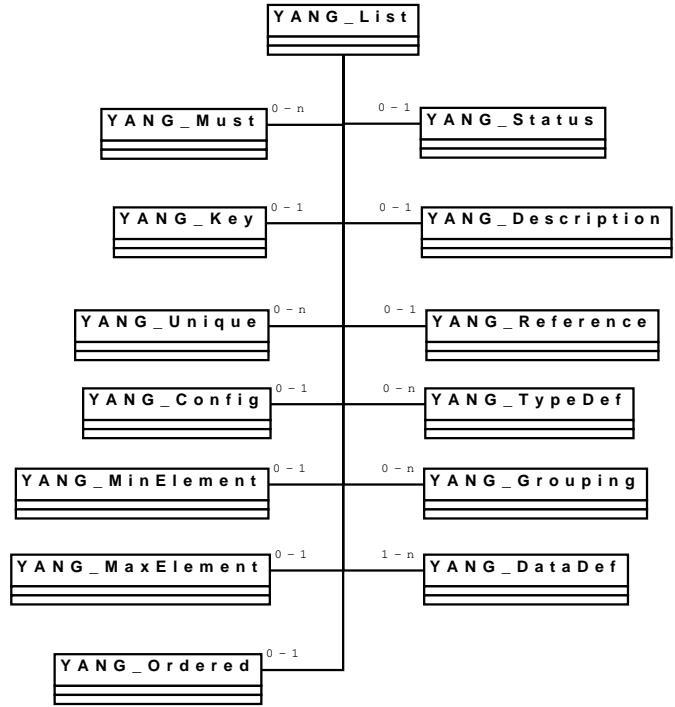}
\end{center}
\caption{List statement classes}
\label{list}
\end{figure}

A {\tt list} statement (fig. \ref{list}) can contain several must,
unique, typedef and grouping statements and must contain at least one
data-def statement.  Key, min element, max element, ordered-by,
status, description and reference are optional.

\subsubsection{Choice statement}
\label{cases:section:global}
\begin{figure}[htbp]
\begin{center}
\includegraphics[scale = .3]{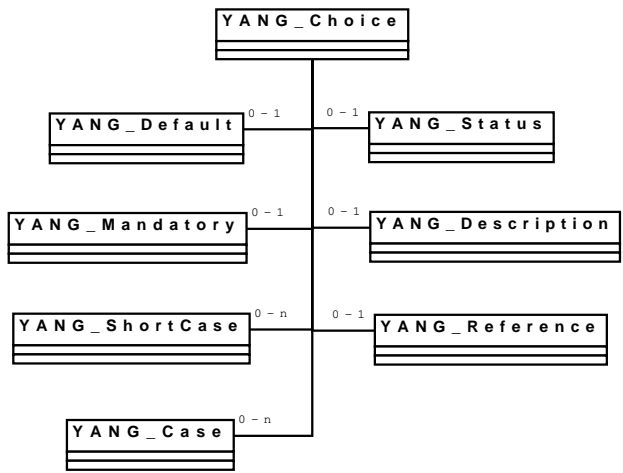}
\end{center}
\caption{Choice statement classes}
\label{choice}
\end{figure}

A  {\tt  choice} statement  (fig.  \ref{choice})  can contain  several
short-case   or  case   statements  that   are  detailed   in  section
\ref{cases:section:detail}.   Default, mandatory,  status, description
and reference are optional.

\subsubsection{Any-xml statement}

\begin{figure}[htbp]
\begin{center}
\includegraphics[scale = .3]{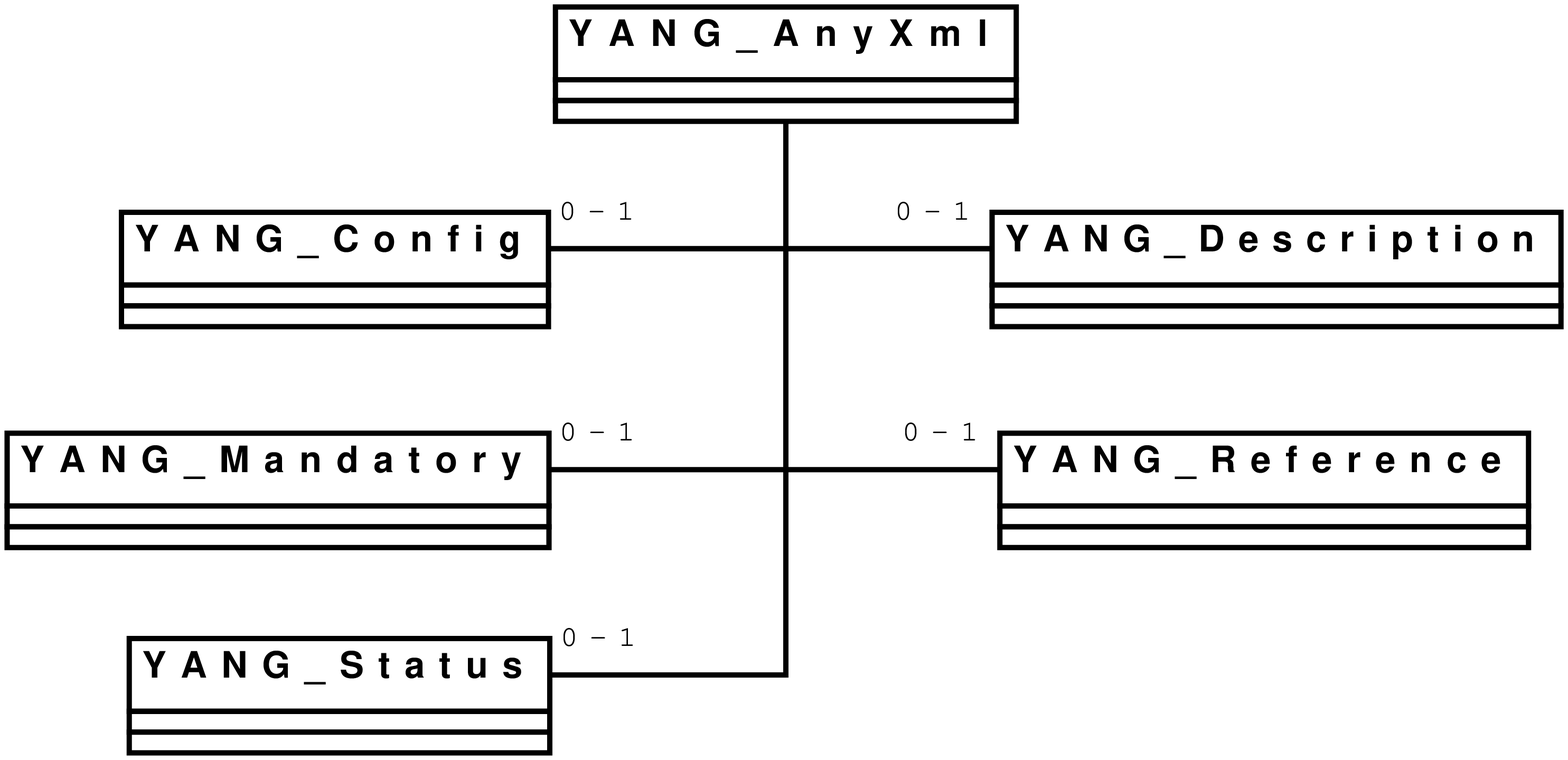}
\end{center}
\caption{Any-xml statement classes}
\label{anyxml}
\end{figure}

An {\tt any-xml} statement (fig.  \ref{anyxml}) can contain a config,
mandatory, status, descrition and reference statements.

\subsubsection{Uses statement}
\label{uses:section:global}

\begin{figure}[htbp]
\begin{center}
\includegraphics[scale = .3]{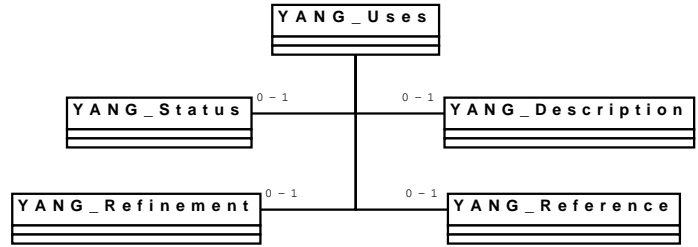}
\end{center}
\caption{Uses statement classes}
\label{uses}
\end{figure}

An  {\tt  uses} statement  (fig.  \ref{uses})  can  contain a  status,
description, reference  and refinement statements.   The refinement is
detailed in section \ref{refinement:section:detail}

\subsubsection{Augment statement}

\begin{figure}[htbp]
\begin{center}
\includegraphics[scale = .3]{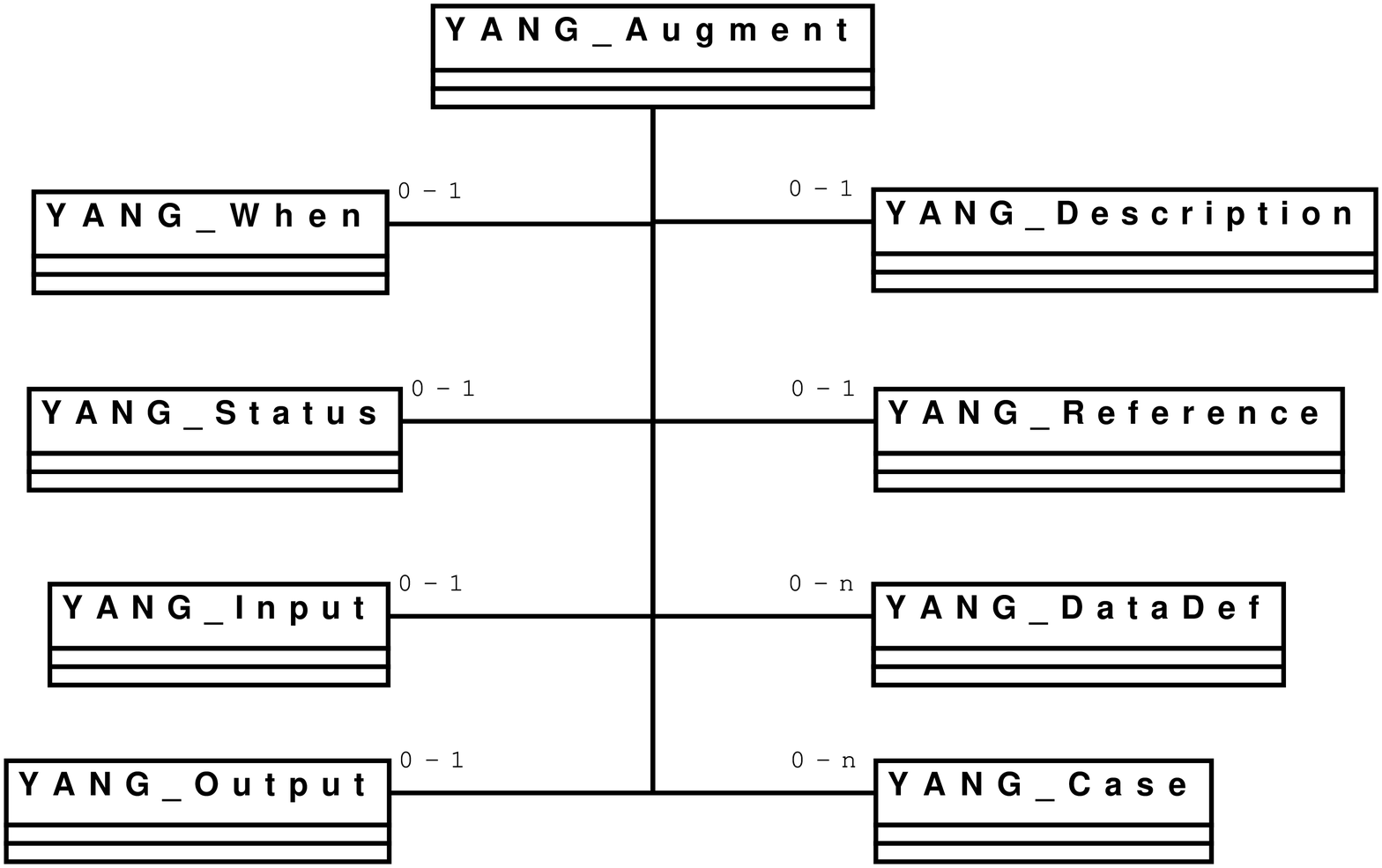}
\end{center}
\caption{Augment statement classes}
\label{augment}
\end{figure}

An {\tt  augment} statement (fig. \ref{augment}) can  contain at least
one datadef or case statements  or one input or output statements.  It
depends on the augmented node. When, status, description and reference
statements are optional.

\subsection{Case and Short Case statements}
\label{cases:section:detail}

Case and short case use are described in section \ref{cases:section:global}.

\subsubsection{Case statement}
\label{casedatadef:section:global}

\begin{figure}[htbp]
\begin{center}
\includegraphics[scale = .3]{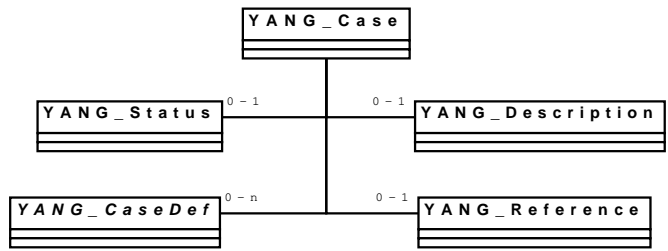}
\end{center}
\caption{Case statement classes}
\label{case}
\end{figure}

A  {\tt  case}  statement   (fig.   \ref{case})  can  contain  several
case-data-def  statements.   Status,  description  and  reference  are
optional.       Case-data-def     is      detailed      in     section
\ref{casedatadef:section:detail}.

\subsubsection{Short Case statement}
\begin{figure}[htbp]
\begin{center}
\includegraphics[scale = .3]{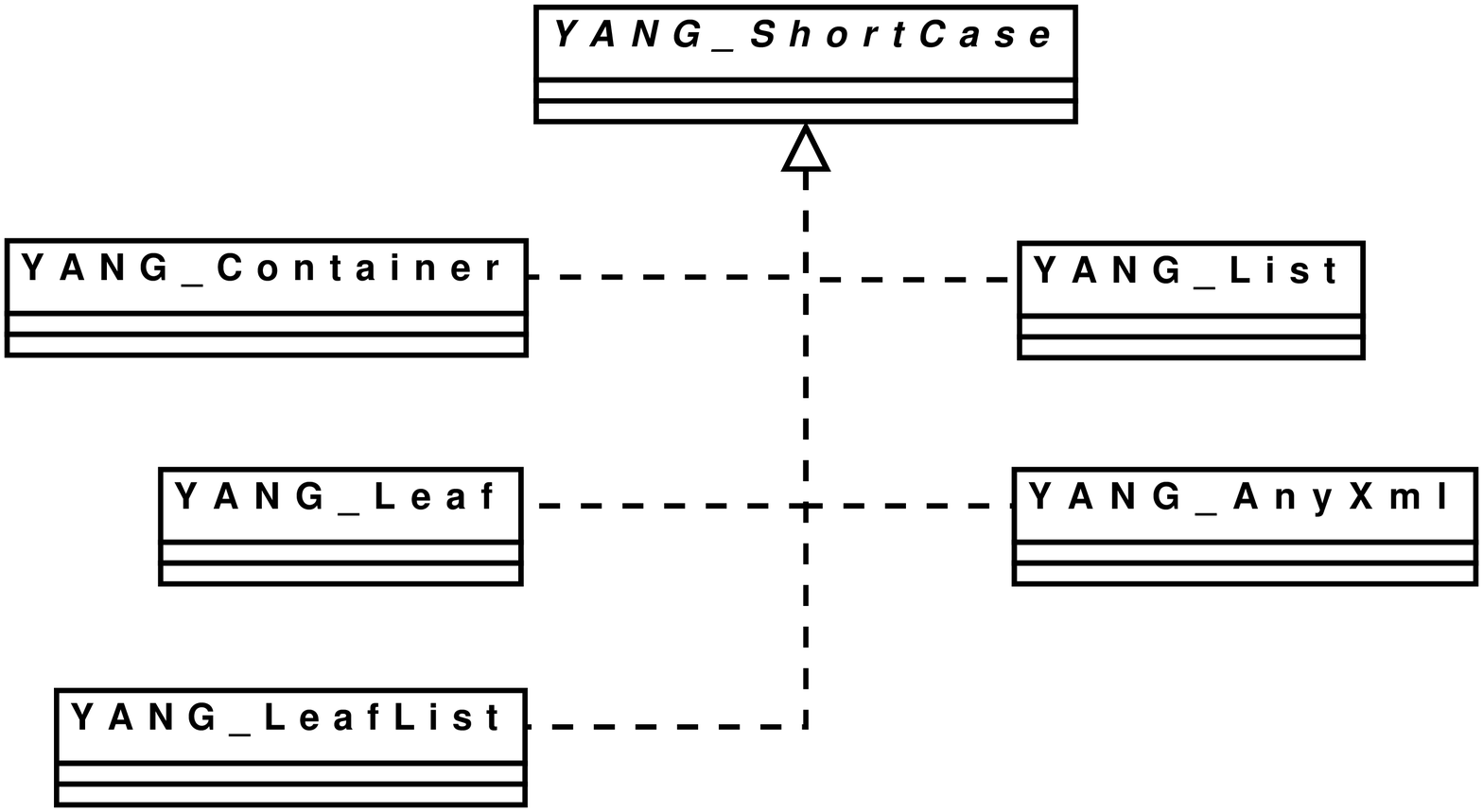}
\end{center}
\caption{Short Case statement classes}
\label{shortcase}
\end{figure}

A short  {\tt case} statement  (fig. \ref{shortcase}) can be  either a
container, leaf, leaf-list, list or any-xml statements.

\subsubsection{Case Data Def statement}
\label{casedatadef:section:detail}

\begin{figure}[htbp]
\begin{center}
\includegraphics[scale = .3]{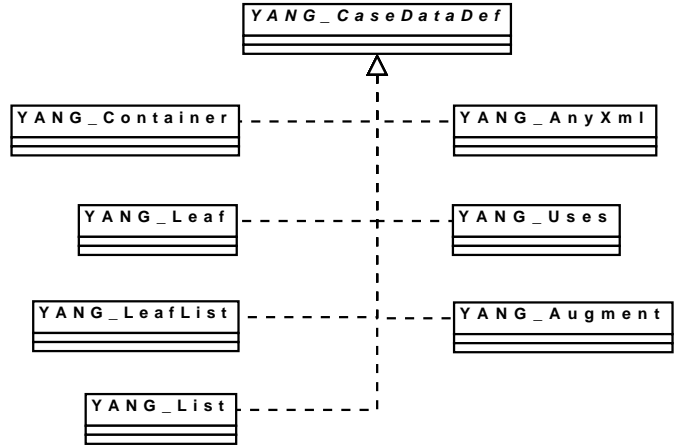}
\end{center}
\caption{Case Data Def  statement classes}
\label{casedatadef}
\end{figure}

A case  data def statement  (fig.  \ref{casedatadef}) can be  either a
container, a  leaf, a  leaf-list, a  list, an any-xml,  an uses  or an
augment  statements.   Case  data  def  use is  described  in  section
\ref{casedatadef:section:global}.

\subsection{Refinement statement}
\label{refinement:section:detail}

\begin{figure}[htbp]
\begin{center}
\includegraphics[scale = .3]{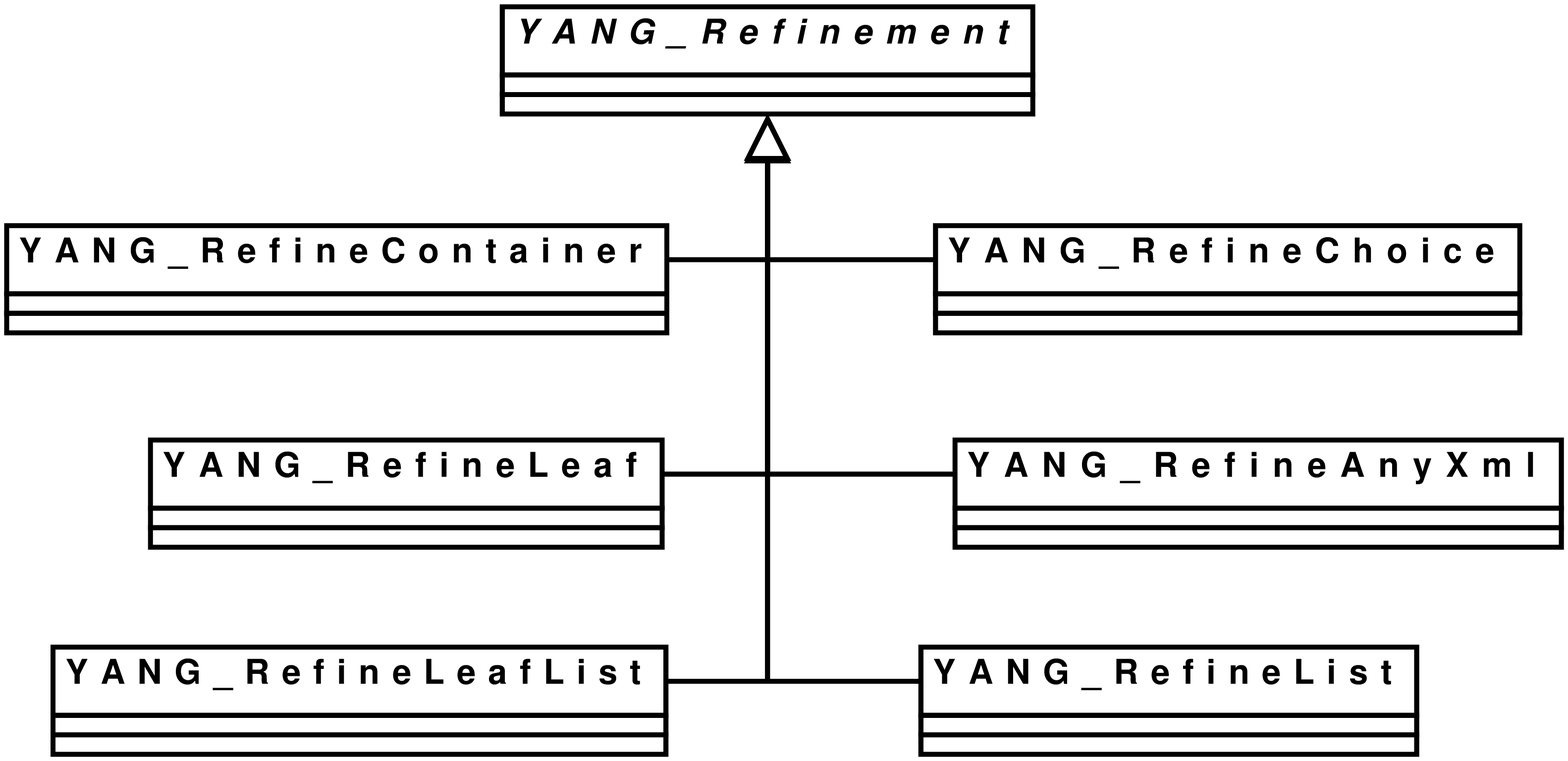}
\end{center}
\caption{Refinement statement classes}
\label{refinement}
\end{figure}

The refinement statement (fig.   \ref{refinement}) can be a refinement
of    a    container,    leaf,    leaf-list,   choice    or    any-xml
statement.     Refinement    use     is    described     in    section
\ref{uses:section:global}.

\subsubsection{Refined Container statement}

\begin{figure}[htbp]
\begin{center}
\includegraphics[scale = .3]{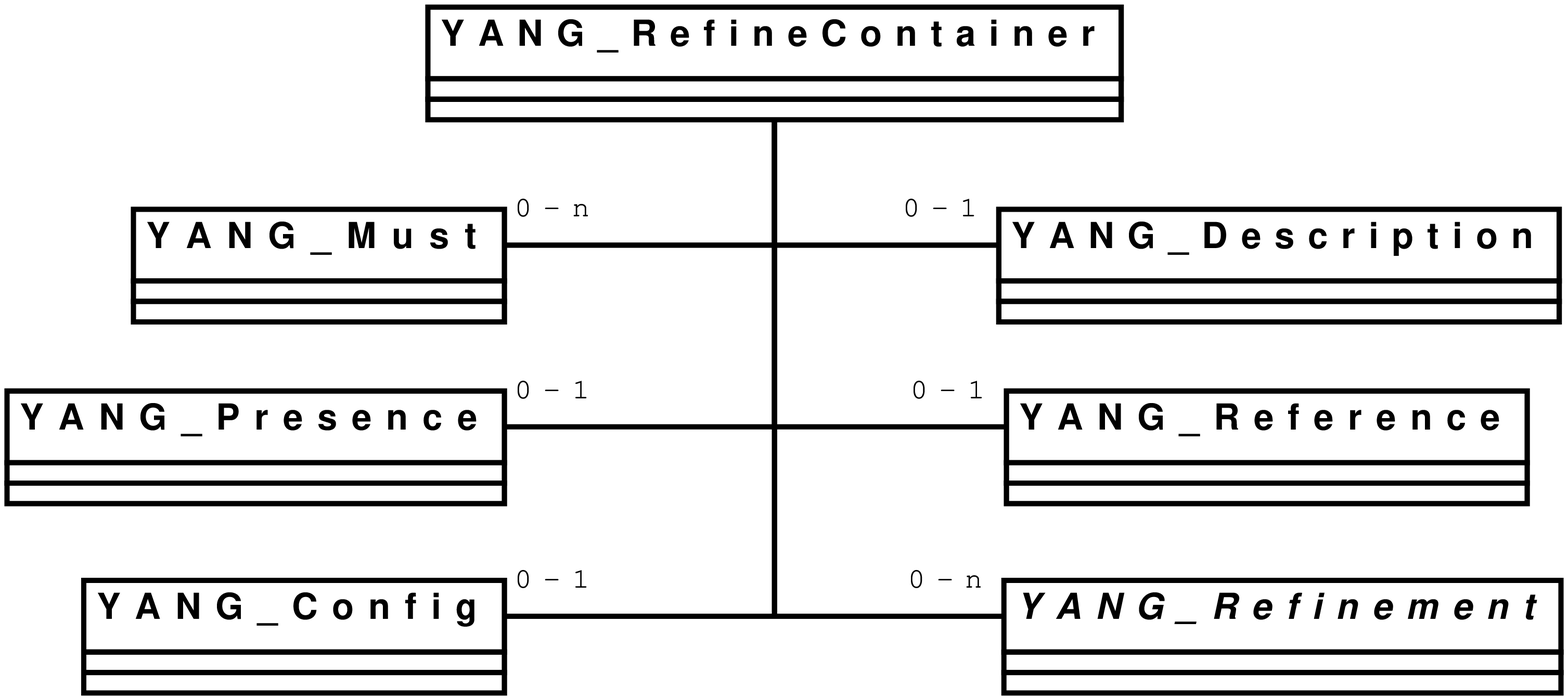}
\end{center}
\caption{Refine Container statement classes}
\label{refinecontainer}
\end{figure}

A refined  {\tt container} statement  (fig. \ref{refinecontainer}) can
contain  several must  and refinement  statements.   Presence, config,
description and reference are optional.

\subsubsection{Refined Leaf statement}

\begin{figure}[htbp]
\begin{center}
\includegraphics[scale = .3]{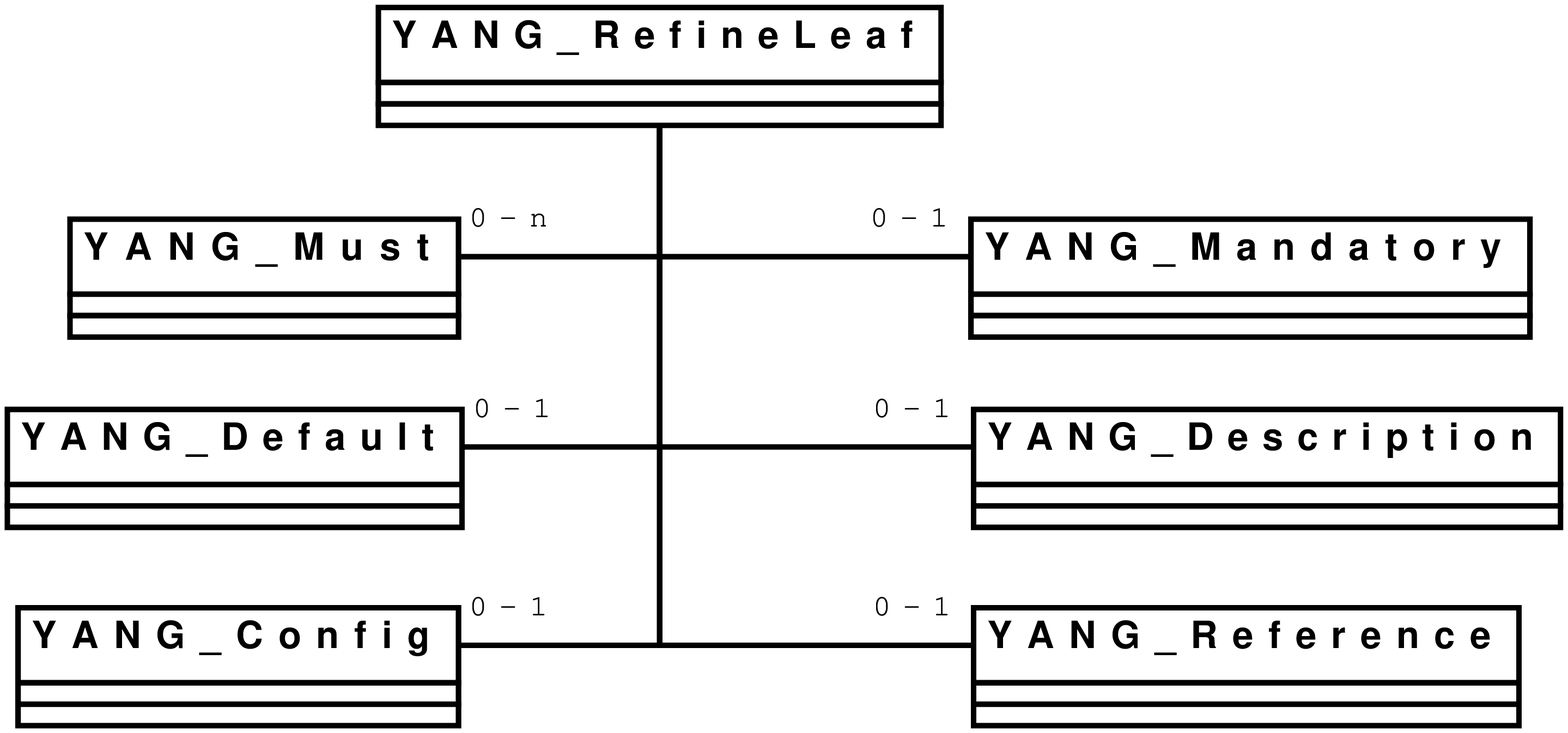}
\end{center}
\caption{Refine Leaf statement classes}
\label{refineleaf}
\end{figure}

A refined  {\tt leaf}  statement (fig.  \ref{refineleaf})  can contain
several must  statements.  Default, config,  description and reference
are optional.

\subsubsection{Refined Leaf List statement}

\begin{figure}[htbp]
\begin{center}
\includegraphics[scale = .3]{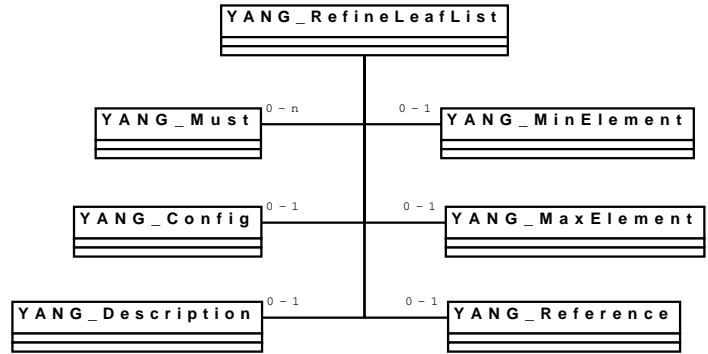}
\end{center}
\caption{Refine Leaf List statement classes}
\label{refineleaflist}
\end{figure}

A  refined {\tt leaf-list}  statement (fig.  \ref{refineleaflist}) can
contain  several must  statements.  Config, min-element,  max-element,
description and reference are optional.

\subsubsection{Refined List statement}

\begin{figure}[htbp]
\begin{center}
\includegraphics[scale = .3]{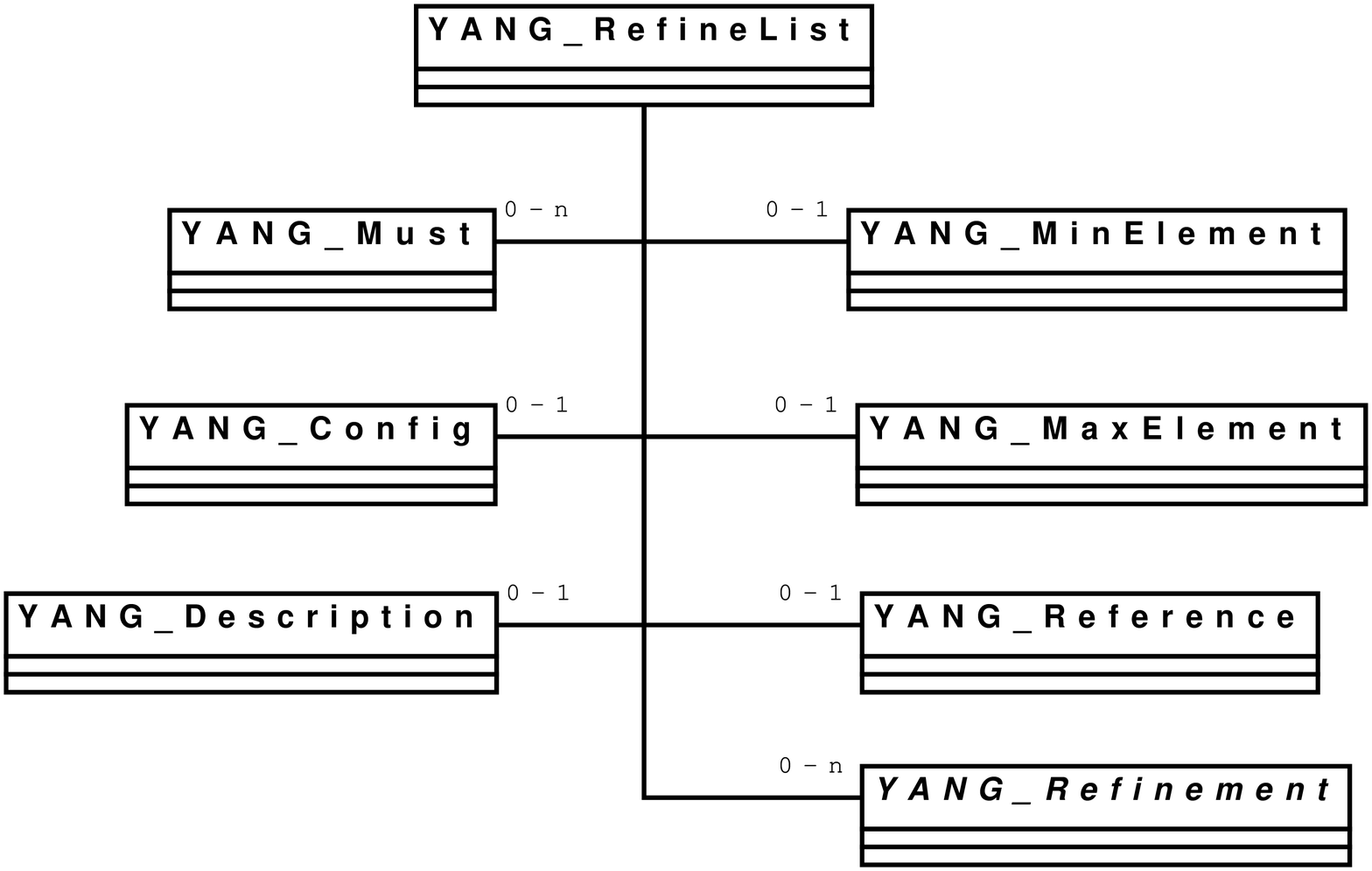}
\end{center}
\caption{Refine List statement classes}
\label{refinelist}
\end{figure}

A refined  {\tt list} statement (fig. \ref{refinelist})  ) can contain
several   must  and   refinement  statements.    Config,  min-element,
max-element, description and reference are optional.

\subsubsection{Refined Choice statement}

\begin{figure}[htbp]
\begin{center}
\includegraphics[scale = .3]{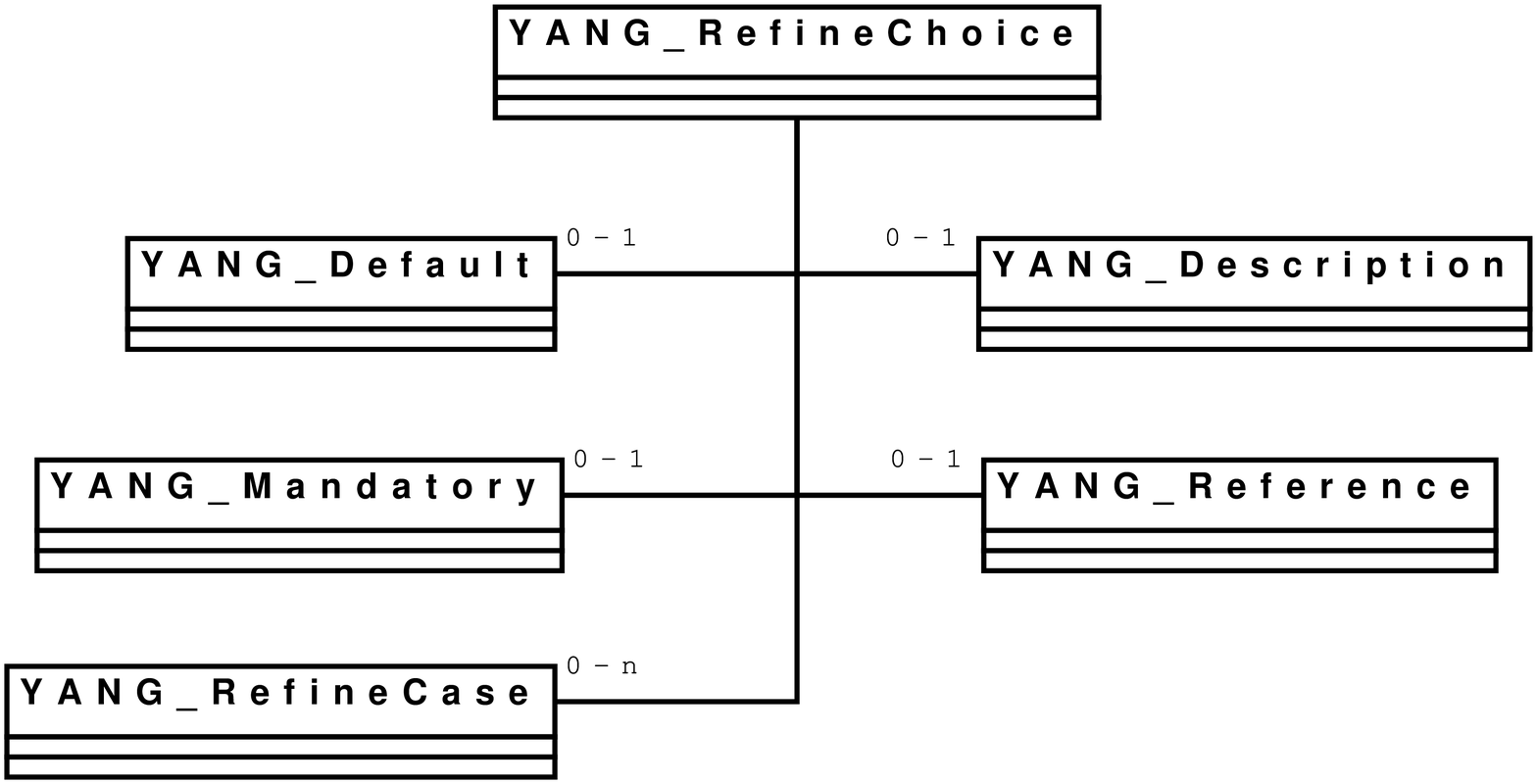}
\end{center}
\caption{Refine Choice statement classes}
\label{refinechoice}
\end{figure}

A refined {\tt case}  statement (fig.  \ref{refinechoice}) can contain
several  refine case statements.  Default, mandatory,  description and
reference are optional.

\subsubsection{Refined Any-xml statement}

\begin{figure}[htbp]
\begin{center}
\includegraphics[scale = .3]{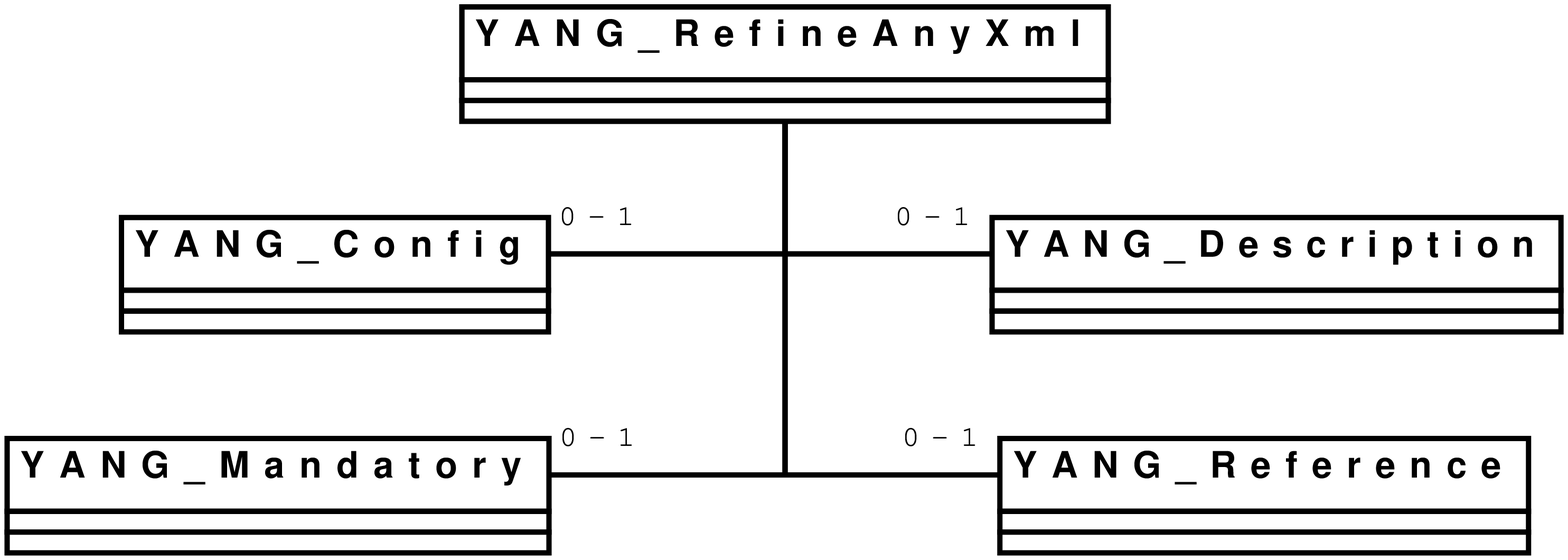}
\end{center}
\caption{Refine Any-xml statement classes}
\label{refineanyxml}
\end{figure}

A refined {\tt any-xml} statement (fig.  \ref{refineanyxml}) optionaly
contains a config, mandatory, description and reference statements.

\subsection{Global view}

\begin{figure}[htbp]
\begin{center}
\includegraphics[scale = 1]{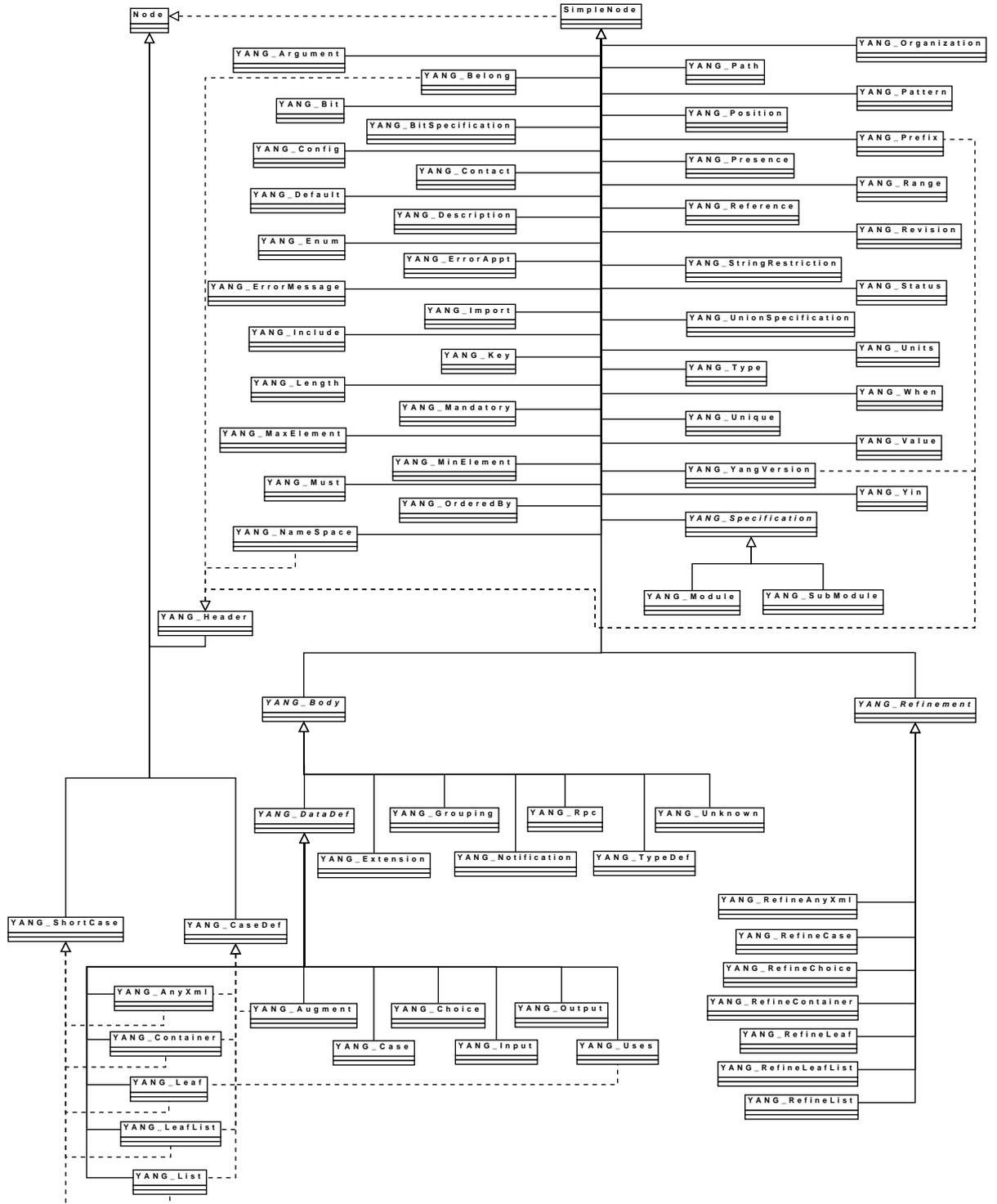}
\end{center}
\caption{YANG Classes and Interfaces}
\label{global}
\end{figure}

The  figure  \ref{global}  shows  all classes  and  their  inheritance
relationships.

\section{Conclusions and future work}

This report describes the \jyang parser and its API. The work is based
on an early release of the draft\cite{yang01}. futhur revisions will
follow the \y\ evolution.

\jyang\ allows  a static parsing  of \y\ specifications but  there are
several other  checks that need to  be done at the  execution time. We
plan to  define some mechanisms to  ensure that a  \nc\ agent realizes
such checks.  The list below may  be not exhaustive but draws our main
goals :
\begin{itemize}
\item
\y\  specifications can  use an  {\tt object-instance}  data  type that
refers to  an existing element in  a configuration. A  \nc\ agent must
verify  that the refered  element  effectively exists,  or  has a  default
value.
\item
\y\ specifications can define new operations and notifications. A \nc\
agent must provide them on top of the RPC mechanism.
\end{itemize}

These evolutions will be bound to a particular \nc\ implementation.

\bibliographystyle{plain}
\bibliography{../../../bibtex/RFC/rfc,../../../bibtex/RFC/draft,../../../bibtex/W3C/w3c}
\end{document}